\documentclass[preprint,authoryear]{elsarticle}

\usepackage{graphicx}
\usepackage{epsfig}
\usepackage{natbib}
\usepackage{colortbl}

\usepackage{changepage}
\usepackage{graphics}
\usepackage{graphicx}
\usepackage{subfigure}
\usepackage{float}

\begin{document}

\title{Phenotypic heterogeneity in modeling cancer evolution}
\author{A. Mahdipour Shirayeh$^{1}$, K. Kaveh$^{2}$, Mohammad Kohandel$^{1, 3}$, Siv Sivaloganathan$^{1,3}$}
\address{$^{1}$Department of Applied Mathematics, University of Waterloo, Waterloo, ON, Canada N2L 3G1\\
$^{2}$Program for Evolutionary Dynamics, Harvard University, Cambridge, 02138 MA, USA\\
$^{3}$Center for Mathematical Medicine, Fields Institute for Research in Mathematical Sciences, Toronto, ON, Canada M5T 3J1}

\begin{abstract}

The unwelcome evolution of malignancy during cancer progression emerges through a selection process in a complex heterogeneous population structure. In the present work, we investigate evolutionary dynamics in a phenotypically heterogeneous population of stem cells (SCs) and their associated progenitors. The fate of a malignant mutation is determined not only by overall stem cell and differentiated cell growth rates but also differentiation and dedifferentiation rates. We investigate the effect of such a complex population structure on the evolution of malignant mutations. We derive exact analytic results for the fixation probability of a mutant arising in each of the subpopulations. The analytic results are in almost perfect agreement with the numerical simulations. Moreover, a condition for evolutionary advantage of a mutant cell versus the wild type population is given in the present study. We also show that microenvironment-induced plasticity in invading mutants leads to more aggressive mutants with higher fixation probability. Our model predicts that decreasing polarity between stem and differentiated cells' turnover would raise the survivability of non-plastic mutants; while it would suppress the development of malignancy for plastic mutants. We discuss our model in the context of colorectal/intestinal cancer (at the epithelium). This novel mathematical framework can be applied more generally to a variety of problems concerning selection in heterogeneous populations, in other contexts such as population genetics, and ecology.
\end{abstract}


\maketitle \vspace{10pt}

\section*{Introduction}

Cancer can be thought of as a complex ecosystem in which not only tumor cells but also other cell types (phenotypes) may influence the overall health of an  organism. Experimental results have recently shown that cancer cells may mimic the functional features of normal cells \citep{JDK}. The most important features are associated with a small subpopulation of cells, namely the stem cells. Stem cells (SCs) are defined to be cells with self-renewal capacity and pluripotency. For instance, they can replenish and regenerate the whole epithelial cell population in normal tissues. It has been proposed that cancer stem cells (CSCs) maintain invasive characteristics, such as (undesirable) multipotency and uncontrolled growth and tumor initiating capacity \citep{Bor,JDO,Rey,Spr,MWC}. The differentiated progenies of SCs are the cells with specialized distinct functions, within the organism. They are produced via a hierarchical division scheme. As the differentiated cells (DCs) become more mature along the hierarchy, their replication potential decreases \citep{ChW,LJ,Mag,RL}.

CSCs reside in small niches and manifest characteristics similar to somatic SCs \citep{Bor}. In solid cancers, CSCs are usually imputed as a result of the expression of similar biomarkers as those used to identify SCs \citep{Gin,Natar,Sheila,Shi,WS}. In colon cancer, the over--expression of the polycomb ring finger oncogene BMI1 leads to the down--regulation of proteins p16INK4a and p14ARF. These proteins override cellular proliferation restriction and generate cancer SLCs \citep{Dim,LJ,MWC}. For mammary stem cells, CD44$^{+}$ and CD24$^{-}$ are reported as markers for stemness. In acute--myeloid leukemia (AML) CD34$^{+}$ CD38$^-$ cells are a leukemia--initiating subpopulation\citep{MWC,RSF}.

Despite the new established dogma that cancer cells originate from a small niche of cells \citep{MWC,RL,Ver}, a range of experiments have now investigated and reported on the cancer initiating capacity of committed progenitor cells \citep{JDO,LiLa,MWC}. In other words, non-SCs can undergo a dedifferentiation process and regain stemness (these cells are also called stem like cells). In breast cancer, epithelial--mesenchymal transition (EMT) factors have been implicated in the production of stem like cells from non--stem cells \citep{mani2008epithelial,chaffer2011normal,chaffer2013poised}. Gupta {\it et al} \citep{Gupta} have also observed that the epithelial differentiated cells with basal markers can convert to cells with stem cell markers (see also \citep{chaffer2011normal}). There are several other experimental observations supporting dedifferentiation of committed progenitor cells \citep{CHH,Hue,LiLa,MWC,Phil}. In addition, this dedifferentiation has been observed, under certain microenvironmental conditions, in normal SCs \citep{JDK,tata2013dedifferentiation,dorantes2012lineage}. In fact, it is becoming apparent that cellular trans--differentiation is activated in a number of organs to produce stem like cells in support of SCs in tissue regeneration \citep{JDK,LiLa,MWC,Ver,ChW,Fau,FK,LJ}.

A variety of quantitative approaches have been utilized to investigate the effect of the stem cell hierarchy and phenotypic heterogeneity on tumor growth \citep{clayton2007single,dhawan2014tumour,turner2009characterization,ZLaP,gao2013acute}.  In the context of cancer evolution, the deterministic population dynamics of the stem cell hierarchy in the absence of plasticity, is discussed by Werner {\it et al} \citep{werner2011dynamics, werner2013deterministic}. Plasticity and dedifferentiation is explored under a diffusion approximation \citep{jilkine2014effect} and replicator equation \citep{kaveh2015replicator}.In \citep{Leili}, the authors discuss the rate of evolution in a simple hierarchical stem and non--stem cell population. They argue that stem cell symmetric division is preferred under natural selection for two-hit mutations.

The evolutionary dynamics of malignant and normal genotypes in the presence of phenotypic transformations (differentiation and dedifferentiation) is not well  understood. In this study, we consider a general framework to study natural selection in heterogeneous populations. We analyze competition between resident and mutant populations which are genotypically different. Each of these types divide into phenotypically different subpopulations (stem cell and differentiated subtypes). Due to homeostasis, the size of SC and DC subpopulations are assumed to remain constant. Stem cells can self--renew and replenish their own population or contribute to differentiated cell population via differentiation events. Differentiated cells can also divide into differentiated cells or dedifferentiate into stem like cell states.

We investigate conditions for the successful selection of a malignant mutation in this complex population structure. Due to the plastic nature of the early malignant progenitors, there is a finite chance for an advantageous mutant to exit the  differentiated group and become part of the SC niche. We derive analytic results that predict the fixation probability of a mutant (either in the SC or DC subpopulations) to establish a finite colony. We assume arbitrary population sizes and division rates and selection intensities as well as (de)differentiation rates. The analytic results are in excellent agreement with stochastic simulations in finite populations. We apply our findings to colorectal cancer and predict dedifferentiation rates that can confer a selection advantage for p53 mutants.

The paper is organized as follows: in the Material and Methods section, the generalized Moran process is defined and the generating function method to calculate long-term fixation or extinction probabilities is presented. The replicator dynamics for the model is derived in the absence of random drift. In the Results section, we discuss the fixation probability as a function of stem cell self-renewal rate, differentiation rate and dedifferentiation rate. The phase diagrams for advantageous mutants are also presented towards the end of this section. In the Discussion section, we summarize our findings and suggest some  possible generalizations to characterize more heterogeneous environments.

\section*{Materials and Methods}

\subsection*{\bf Generalized Moran process with (a)symmetric division and plasticity}

Consider two populations of resident or wild type (type 1) and mutant or invader (type 2). Mutants are the result of an oncogenic mutation in the resident population. Each genotype is divided into phenotypically different subpopulations of stem cells (SC) and differentiated cells (DC). Stem cells can self-renew symmetrically where the offspring are stem cells. They can differentiate (symmetrically or asymmetrically) to produce differentiated progenies of the same genotype. We denote the probability of asymmetric differentiation (per division) by $\hat{u}_{1},\hat{u}_{2}$ and symmetric differentiation by $\hat{v}_{1},\hat{v}_{2}$. The overall probability of differentiation is $u_{1,2} = \hat{u}_{1,2} + 2\hat{v}_{1,2}$. This is due to the fact that symmetric differentiation produces two differentiated cells. Similarly the self-renewal probability is denoted by $1-u_{1,2}$. The indexes 1 or 2 denote the corresponding probabilities for a wild type or mutant. The division rate of a normal (or mutant) stem cell is denoted by $r_{1}$ ($r_{2}$) respectively. Similarly, the division rates of progenitors/differentiated cells are denoted by $\hat{r}_{1}$ and $\hat{r}_{2}$.
\begin{figure}[h!]
\centering
\includegraphics[angle=270,width=0.95\textwidth]{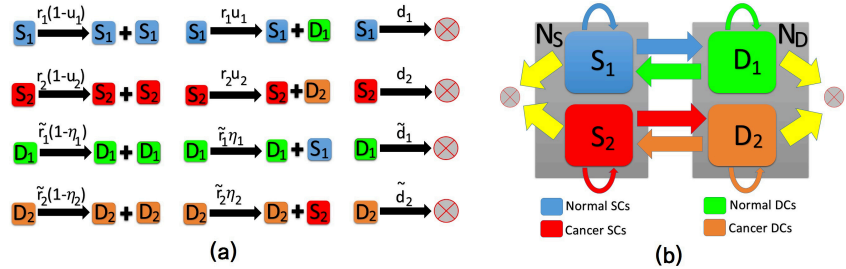}\\[-2.5cm]
\caption { {\bf Phenotypic--genotypic changes in individuals within a four-- compartmental structure.} We consider constant population sizes $N_{\rm S}$ and $N_{\rm D}$ for SCs and DCs respectively. With respect to the finite Markov chain, we consider a generalized model to take into account the competition between normal and malignant individuals in each of the SC and DC subpopulations. Differentiation and dedifferentiation events connect the selection dynamics between the two niches. In (a), all possible differentiation, dedifferentiation, and death events with their corresponding rates are represented. The SC-DC compartmental structure is depicted in (b) with the associated self--renewal and differentiation/plasticity possibilities.}
\label{fig888}
\end{figure}
For evolutionary dynamics we consider a birth-death (BD) Moran process as follows: At each time step, an individual is chosen to reproduce proportional to its fitness within the SC or DC compartments. If a normal (mutant) cell in the SC compartment is chosen to reproduce, its offspring replaces a randomly chosen cell in the stem cell compartment with probability $1-u_{1}$ ($1-u_{2}$). Otherwise, with probability  $u_{1}$ ($u_{2}$), the (differentiated) offspring replaces a randomly chosen cell in the DC compartment. Similarly, if a differentiated cell is chosen to reproduce, its offspring replaces another cell in the differentiated cell compartment with the probability $1-\eta$. Alternatively, the offspring can dedifferentiate into a stem-like cell and replace a randomly chosen individual in the stem cell compartment with a rate $\eta$, where $\eta = \eta_{1} (\eta = \eta_{2}$) denote the dedifferentiation probability for normal (mutant) DCs. For simplicity we assumed death rates of all types to be equal and set this to unity.

\begin{table}[h!]
\centering{
   \caption{{\bf Model parameters.}} \label{table1}
    \begin{tabular}{l l }
        \hline
        \rowcolor[rgb]{0.8,0.8,1} Notation & Description \\ \hline
      $N_{\rm S},N_{\rm D}$ & Total number of stem and differentiated cells \\
      $r_{\rm 1},r_{\rm 2} $ & Net reproduction rate of wild type and mutant stem cells\\
      $\tilde{r}_{\rm 1},\tilde{r}_{\rm 2} $ & Net reproduction rate of  wild type and mutant differentiated cells  \\
      $u_1, u_2$ & Asymmetric differentiation rate of normal and mutant stem cells \\
      $\eta_1, \eta_2$ & Dedifferentiation rate of normal and mutant differentiated cells\\
      \hline
\end{tabular}
}
\end{table}
The above dynamics models the differentiation mechanism with an {\it effective} asymmetric division with the probabilities $u_{1},u_{2}$. Thus in the following we use the terms differentiation (of stem cells) and asymmetric division interchangeably.

The above Moran process can be written as a continuous time process ($1/N$ is the duration of each time step for $N=N_{\rm S}+N_{\rm D}$)
\begin{eqnarray}
\frac{1}{N}\frac{\partial p(n_{\rm S},n_{\rm D};t)}{\partial t} &=&  W^{+}_{{\rm S}}(n_{\rm S}-1,n_{\rm D})\, p(n_{\rm S}-1,n_{\rm D};t) +   W^{-}_{{\rm S}}(n_{\rm S}+1,n_{\rm D})\,p(n_{\rm S}+1,n_{\rm D};t)\nonumber\\
&& \hspace{-2cm} +\, W^{+}_{{\rm D}}(n_{\rm S},n_{\rm D}-1)\, p(n_{\rm S},n_{\rm D}-1;t) +   W^{-}_{{\rm D}}(n_{\rm S},n_{\rm D}+1)\, p(n_{\rm S},n_{\rm D}+1;t)   \nonumber\\
&& \hspace{-2cm} -\,\big( W^{+}_{{\rm S}}(n_{\rm S},n_{\rm D})+W^{+}_{{\rm D}}(n_{\rm S},n_{\rm D})+W^{-}_{{\rm S}}(n_{\rm S},n_{\rm D})+W^{-}_{{\rm D}}(n_{\rm S},n_{\rm D})\big)\, p(n_{\rm S},n_{\rm D};t). \nonumber\\
\label{eq:master}
\end{eqnarray}
where $p(n_{\rm S}, n_{\rm D}; t)$ denotes the probability of having $n_{\rm S}$ mutant stem cells and $n_{\rm D}$ mutant differentiated cells, at time $t$ (given $n^0_{\rm S}$ and $n^0_{\rm D}$ at $t=0$). The population of normal cells are given by $N_{\rm S}-n_{\rm S}$ and $N_{\rm D}-n_{\rm D}$ correspondingly. The probabilities $W^{\pm}_{\rm S}$  and $W^{\pm}_{\rm D}$ are the transition probabilities corresponding to an increase or decrease by one in the number of mutant SCs and DCs resp. They are given by
\begin{eqnarray}
W^{+}_{{\rm S}}(n_{\rm S},n_{\rm D}) &=& {\rm Prob}(n_{\rm S},n_{\rm D} \rightarrow n_{\rm S}+1,n_{\rm D})\nonumber\\ &=& \left( \frac{r_{2}\,(1-u_{2})\,n_{\rm S} +  \tilde{r}_{2}\,\eta_{2}\, n_{\rm D}}{  {\rm N}_{{\rm r}}  } \right) \frac{N_{\rm S}-n_{\rm S}}{N_{\rm S}}, \nonumber\\
W^{-}_{{\rm S}}(n_{\rm S},n_{\rm D})  &=& {\rm Prob}(n_{\rm S},n_{\rm D} \rightarrow n_{\rm S}-1,n_{\rm D})\nonumber\\ &=& \left( \frac{r_{1}\, (1-u_{1})\, (N_{\rm S}-n_{\rm S})  + \tilde{r}_{1}\, \eta_{1}\, (N_{\rm D}-n_{\rm D})}{    {\rm N}_{{\rm r}}}  \right) \, \frac{n_{\rm S}}{N_{\rm S}}, \nonumber\\[-3mm]
&& \label{eq9} \\[-2mm]
W^{+}_{{\rm D}}(n_{\rm S},n_{\rm D})  &=& {\rm Prob}(n_{\rm S},n_{\rm D} \rightarrow n_{\rm S},n_{\rm D}+1) \nonumber\\&=& \left( \frac{ \tilde{r}_{2}\, (1-\eta_{2})\, n_{\rm D}+  r_{2}\, u_{2}\, n_{\rm S}}{   {\rm N}_{{\rm r} }} \right)\, \frac{N_{\rm D}-n_{\rm D}}{N_{\rm D}}, \nonumber\\
W^{-}_{{\rm D}} (n_{\rm S},n_{\rm D}) &=& {\rm Prob}(n_{\rm S},n_{\rm D} \rightarrow n_{\rm S},n_{\rm D}-1) \nonumber\\ &=& \left( \frac{ \tilde{r}_{1}\, (1-\eta_{1})\, (N_{\rm D}- n_{\rm D})  +   r_{1}\, u_{1}\, (N_{\rm S}-n_{\rm S}) }{   {\rm N}_{{\rm r}}} \right)\, \frac{n_{\rm D}}{N_{\rm D}}. \nonumber
\label{eq:transition}
\end{eqnarray}

The denominator $ {\rm N}_{{\rm r}}$ denotes the total fitness of SC and DC individuals:
\begin{eqnarray}
 {\rm N}_{{\rm r}} = r_{1}\,(N_{\rm S}-n_{\rm S})  + r_{2}\, n_{\rm S} + \tilde{r}_{1}(N_{\rm D}-n_{\rm D}) +  \tilde{r}_{2}\, n_{\rm D}.
\end{eqnarray}
The above Markov process has two absorbing states corresponding to fixation or extinction of the mutant or WT. The competition between the two genotypes in the stem cell compartment is tied to the competition inside the differentiated compartment via differentiation and dedifferentiation mechanisms. In the absence of plasticity we have a hierarchical population structure where only mutations in the stem cell compartments can give rise to fixation in the whole population.

In the next section,  we present analytic solutions for the probability of fixation, of an invading mutant as a function of division rates and (de)differentiation rates.
Our calculations match with simulation results for a wide range of parameters and population sizes.

\subsection*{\bf Fixation probability in a heterogeneous Moran process}

One of the most important questions to address within a heterogeneous population is the chance of success for a mutation in different subtypes.

The fixation probability of a mutant originating in the stem cell compartment, $\rho_{\rm S}$ or the differentiated cell compartment, $\rho_{\rm D}$ is a measure of the tumor initiating capacity of each subpopulation. For a completely hierarchical population, only mutants that arise in the stem cell niche have a chance of fixating in the whole population thus $\rho_{\rm S} = \rho$. If the progenitors can dedifferentiate into stem-like cells, the comparison between the two fixation probabilities, ($\rho_{\rm S}$ and $\rho_{\rm D}$), is a good measure of how the tumor initiating capacity correlates with the notion of stemness.

The use of the probability generating function (PGF) method to study a constant population Moran process is discussed in \citep{HV, HV1,kaveh2015duality}. It is used to present an alternative derivation of the (well-mixed) Moran fixation probability, by identifying a martingale for the process. Here we generalize this technique for a heterogeneous population under selective pressure in the presence of phenotypic plasticity.

A martingale for the above four population model, Eqs. (\ref{eq:master}), (\ref{eq:transition}), can be written as
\begin{equation}
\langle \big(z_{\rm S}^{\star}\big)^{n_{\rm S}}\big(z_{\rm D}^{\star}\big)^{n_{\rm D}}\rangle,
\end{equation}
where $\langle \cdot \rangle$ denotes the stochastic average and the (auxiliary) variables $z^{\star}_{\rm S}$ and $z^{\star}_{\rm D}$ satisfy the following system of algebraic equations

\begin{eqnarray}
&& (z^{\star}_{\rm S}-1)\left[  r_2(1-u_2) \, z^{\star}_{\rm S}- r_1(1-u_1)-\tilde{r}_1\, \eta_1 \right] + (z^{\star}_{\rm D}-1)\, z^{\star}_{\rm S}\, r_2\,u_2 =0 \nonumber\\[-3mm]
&& \label{eqbd}  \\[-2mm]
&& (z^{\star}_{\rm D}-1)\left[ \tilde{r}_2\, (1-\eta_2 )\, z^{\star}_{\rm D} - \tilde{r}_1\,(1-\eta_1) - r_1\, u_1 \right] + (z^{\star}_{\rm S}-1)\, z^{\star}_{\rm D} \, \tilde{r}_2\, \eta_2=0. \nonumber
\end{eqnarray}

By matching the initial conditions for $t=0$ and the steady state solutions of the PGF, we can obtain analytic expressions for the fixation probability of mutants of each subtype (stem or differentiated). In general, the fixation probability beginning with $i$ mutant SCs and $j$ mutant progenitors is (See SI for details)
\begin{eqnarray}
\rho_{ij}= \frac{1- \left(z_{\rm S}^{\star}  \right)^i \,   \left(z_{\rm D}^{\star}  \right)^j }{  1- \left(z_{\rm S}^{\star}  \right)^{N_{\rm S}} \,\left(z_{\rm D}^{\star}  \right)^{N_{\rm D}} }. \label{FP}
\end{eqnarray}
For $i=1$ and $j=0$ (starting with one initial SC mutant) the fixation probability is
\begin{eqnarray}
\rho_{\rm S}\equiv\rho_{10}=  \frac{1- z_{\rm S}^{\star} }{  1- \left(z_{\rm S}^{\star}  \right)^{N_{\rm S}}  \left(z_{\rm D}^{\star}  \right)^{N_{\rm D}} }. \label{FP1}
\end{eqnarray}
Similarly, the fixation probability of a newborn mutant in the DC compartment ($i=0,j=1$) is
\begin{eqnarray}
\rho_{\rm D}\equiv\rho_{01}=  \frac{1- z_{\rm D}^{\star} }{  1- \left(z_{\rm S}^{\star}  \right)^{N_{\rm S}}  \left(z_{\rm D}^{\star}  \right)^{N_{\rm D}} }. \label{FP2}
\end{eqnarray}
Moreover, assuming random mutations, i.e. uniform mutation rates in both compartments, the average fixation probability is given by
\begin{eqnarray}
\rho = \frac{1 - (N_{\rm S} /N_{\rm tot})z^{\star}_{\rm S} - (N_{\rm D} /N_{\rm tot})z^{\star}_{\rm D}}{  1- \left(z_{\rm S}^{\star}  \right)^{N_{\rm S}}  \left(z_{\rm D}^{\star}  \right)^{N_{\rm D}} }. \label{FP3}
\end{eqnarray}
with $N_{\rm tot} = N_{\rm S}+N_{\rm D}$. The probability of a successful emergent mutant {\it before} time $t$ (from a background of $N_{\rm tot}$ normal cells) is given by
\begin{eqnarray}
P(t) = 1 - e^{-N_{\rm tot}\cdot \mu\rho t},
\end{eqnarray}
\noindent where $\mu$ denotes the mutation rate.
\subsection*{\bf Stochastic simulation}

Using the model described above, we performed numerical simulations using such updates until each of the runs tends to saturation in the fraction of SCs and DCs, or until we reach the maximum updating time of  T=15,000 for each realization. Then running the whole procedure for 20,000 realizations, we calculated the fraction of results for the fixation probability of SCs and DCs in those runs. Then repeating each calculation for a set of five iterations, we calculated the mean and error bars. Errors are calculated as the standard deviation of the mean.

\subsection*{\bf Replicator equation}

The Markov process considered in our four--compartment model exhibits deterministic dynamics in the absence of stochastic fluctuations, i.e. infinite population limit. This replicator equation captures the average frequency of various phenotypes which provides insight into the evolutionary dynamics of the system. Firstly, we detect the frequency of each phenotype at the fixed points and secondly, analyze the phase diagram by varying different parameter values (see the Results Sec.) at equilibrium. Starting from the master equation (\ref{eq:master}), one obtains the following system of deterministic equations for malignant SCs and DCs (see SI for derivation)
\begin{eqnarray}
\frac{d x_{\rm S}}{dt}  &=&  \frac{   [ r_2\,(1- u_2) - r_1\,(1-u_1)  ] \,x_{\rm S}\,(1- x_{\rm S})  + \tilde{r}_2\,\eta_2\,x_{\rm D}\,(1- x_{\rm S})  - \tilde{r}_1\eta_1\,x_{\rm S}\,(1- x_{\rm D})    } {  r_1\, (1-x_{\rm S}) + r_2\,  x_{\rm S} + \tilde{r}_1 \, (1-x_{\rm D}) +\tilde{r}_2 \, x_{\rm D} },  \nonumber \\
\frac{d x_{\rm D}}{dt}  &=& \frac{   [ \tilde{r}_2\,(1- \eta_2) - \tilde{r}_1\,(1- \eta_1)  ] \,x_{\rm D}\,(1- x_{\rm D})  + r_2\,u_2\,x_{\rm S}\,(1- x_{\rm D})  - r_1\, u_1\,x_{\rm D}\,(1- x_{\rm S})    } {  r_1\, (1-x_{\rm S}) + r_2\,  x_{\rm S} + \tilde{r}_1 \, (1-x_{\rm D}) +\tilde{r}_2 \, x_{\rm D} },\nonumber\\
 \label{eq:RepM4}
\end{eqnarray}
where $x_{\rm S} = \langle n_{\rm S}(t)/N_{\rm S} \rangle$ and $x_{\rm D} = \langle n_{\rm D}/N_{\rm D} \rangle $ are the average frequencies of mutant SCs and DCs resp. Stationary state frequencies, $x^{\star}_{\rm S}$ and $x^{\star}_{\rm D}$ satisfy the following coupled system of equations
\begin{eqnarray}
&&\hspace{-1.25cm} [ r_2\,(1- u_2) - r_1\,(1-u_1)  ] \,x^{\star}_{\rm S}\,(1- x^{\star}_{\rm S})  + \tilde{r}_2\,\eta_2\,x^{\star}_{\rm D}\,(1- x^{\star}_{\rm S})  - \tilde{r}_1\eta_1\,x^{\star}_{\rm S}\,(1- x^{\star}_{\rm D}) = 0,  \nonumber \\[-2mm]
\nonumber\\
&&\hspace{-1.25cm} [ \tilde{r}_2\,(1- \eta_2) - \tilde{r}_1\,(1- \eta_1)  ] \,x^{\star}_{\rm D}\,(1- x^{\star}_{\rm D})  + r_2\,u_2\,x^{\star}_{\rm S}\,(1- x^{\star}_{\rm D})  - r_1\, u_1\,x^{\star}_{\rm D}\,(1- x^{\star}_{\rm S}) = 0.\nonumber\\
\label{eqRD}
\end{eqnarray}
In the next section, after obtaining the exact analytic form of the fixation probability for this generalized Moran process, we will use Eqs. (\ref{eqRD}) to derive the condition for evolutionary advantage of the mutant genotype and the phase diagram for the model.

\section*{Results}

\subsection*{\bf Exact analytic results for the fixation probability}

We first begin with the analytic expressions for the  survival probabilities Eqs. (\ref{FP})-(\ref{FP3}) where $z^{\star}_{\rm S}$ and $z^{\star}_{\rm D}$ are the solutions of Eqs. (\ref{eqbd}). We begin with some simpler limiting cases where compact algebraic results can be obtained, and then proceed to general solutions of Eqs. (\ref{FP})-(\ref{FP3}).

\paragraph{Standard Moran process.}
Let us consider the simple case where the differentiation and dedifferentiation rates are set to zero. In this case our model should reduce to two disjoint Moran processes, one for the stem cell and one for the differentiated cell compartments.

For this case, we obtain $z^{\star}_{\rm S}=r_{1}/r_{2}$ and $z^{\star}_{\rm D}=\tilde{r}_{1}/\tilde{r}_{2}$ and the fixation probability for a mutant to dominate a SC (DC) niche is
\begin{eqnarray}
\rho_{1}= \displaystyle{ \frac{1- \frac{r_1}{r_2} }{  1- \left(\frac{r_1}{r_2}  \right)^{N_{\rm S}}   }  },\hspace{1cm}
\rho_{2}=  \displaystyle{\frac{1- \frac{\tilde{r}_1}{ \tilde{r}_2} }{  1- \left(\frac{\tilde{r}_1}{ \tilde{r}_2}  \right)^{N_{\rm D}} }    }. \label{Noplast}
\end{eqnarray}
Another interesting limit that resembles a well-mixed Moran process in the stem cell compartment occurs when $\tilde{r}_{1,2} \simeq 0$ but $u_{1,2}\neq 0$. This is an exaggerated case showcasing the limited proliferation capacity of differentiated progeny. The corresponding average fixation probability for an emerged mutant in the SC or DC compartment is
\begin{eqnarray}
\rho= {\displaystyle{ \frac{ N_{\rm S}\left(1-  \frac{r_1}{r_2}\right) }{ N_{\rm tot}\left(  1- \left(\frac{r_1}{r_2}  \right)^{N_{\rm S}}  \right) }  }},
\end{eqnarray}
since the fixation probability of a newborn mutant in the DC class will be zero in this case.
\begin{figure}[h!]
 \centering
 \subfigure[ ]{\includegraphics[width=0.47\textwidth]{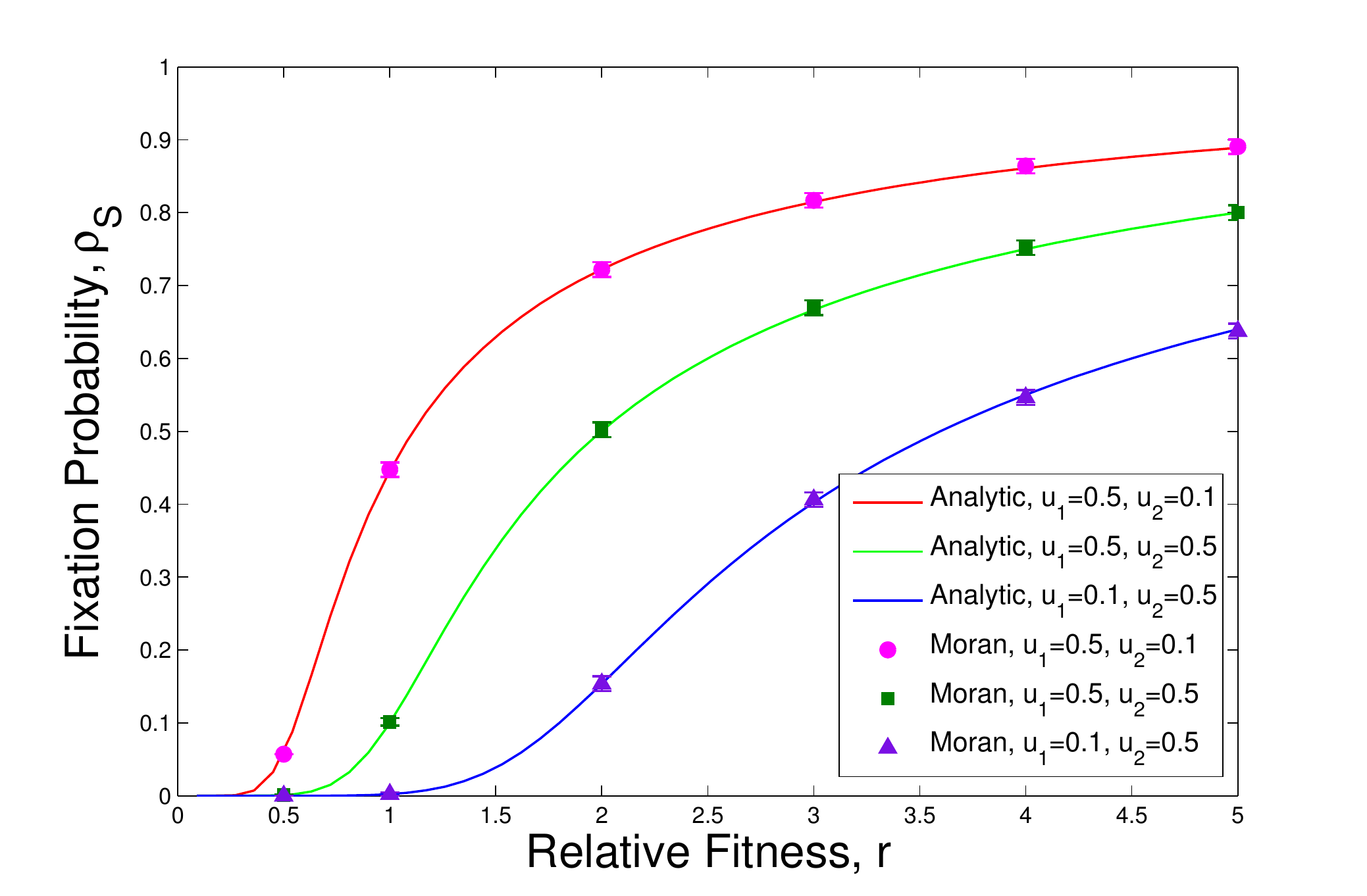} }
\subfigure[ ]{\includegraphics[width=0.47\textwidth]{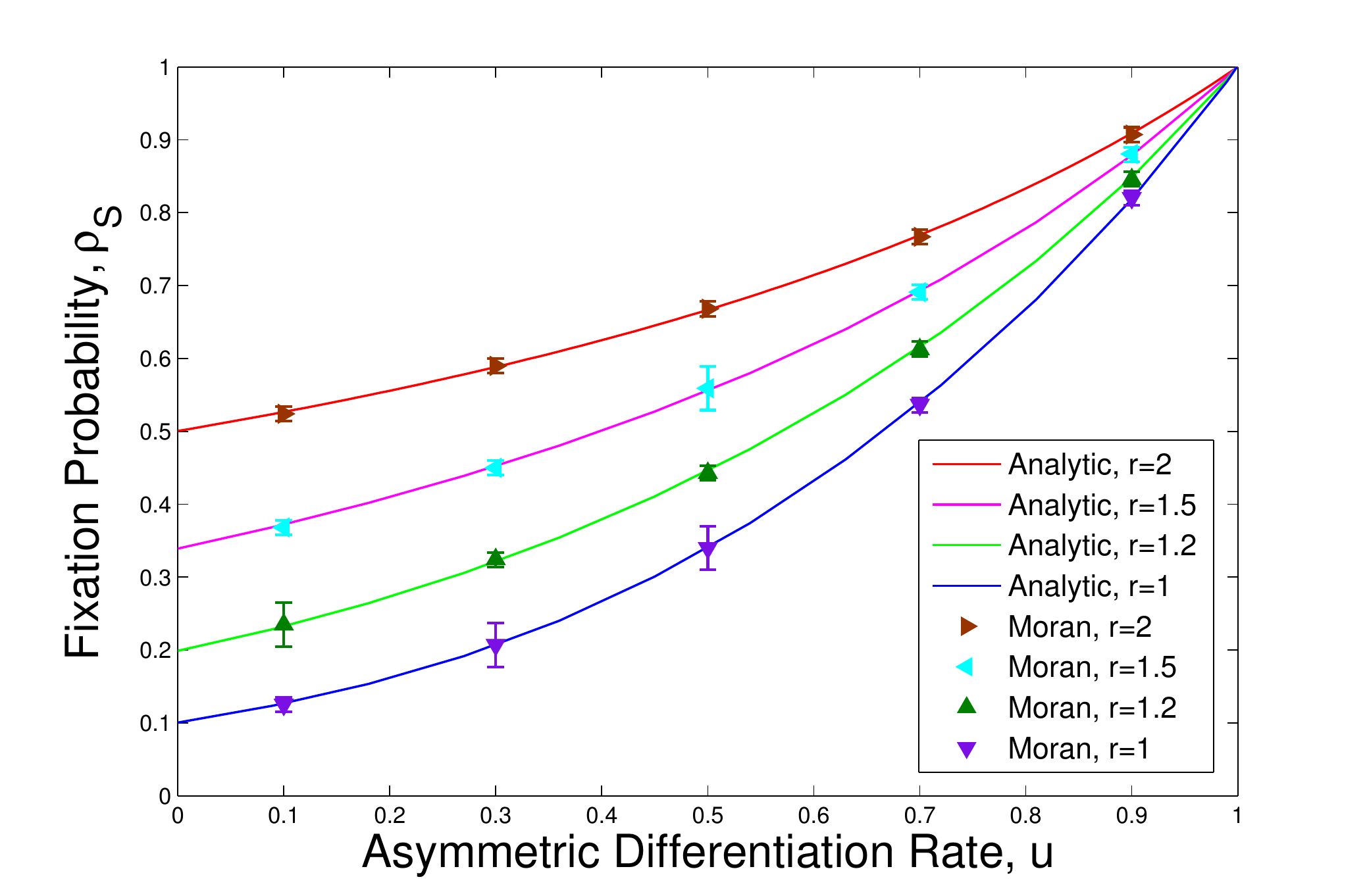}}\\[-2mm]
\subfigure[ ]{\includegraphics[width=0.47\textwidth]{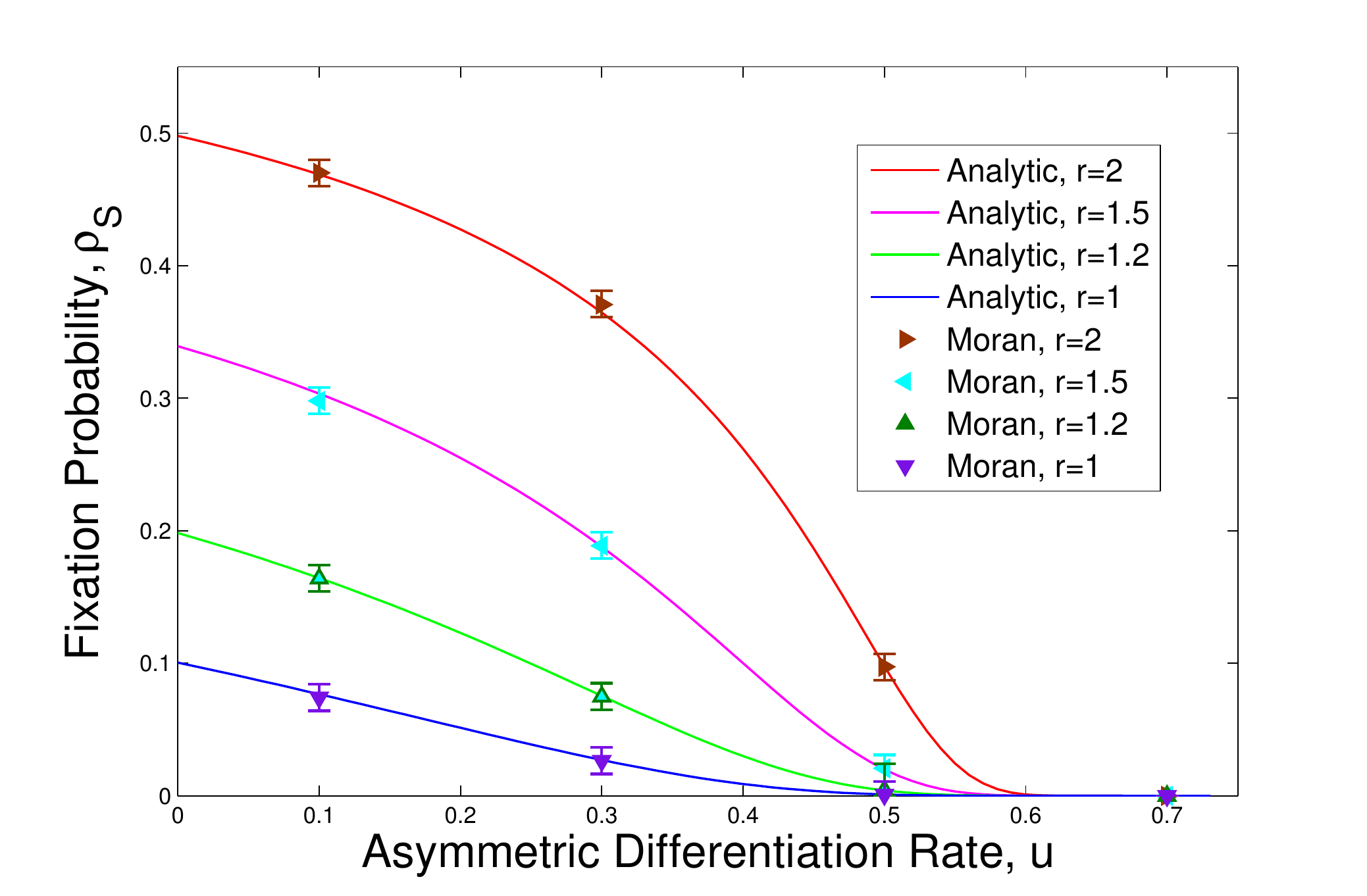}}
\caption { {\bf The fixation probability of mutants in the absence of plasticity.} We assume $N_{\rm S}=N_{\rm D}=10$, $r_1=\tilde{r}_1=1$, and $\eta_1=\eta_2= 0$ in simulations (points) and analytic calculations (solid lines). Each error bar shown at each point is the standard deviation of mean. In (a) changing parameters $u_1, u_2$, that are the differentiation rates of normal and tumor SCs resp., the trends for the fixation probability of mutants is given as a function of relative fitness of mutants, referred to as $r_2=\tilde{r}_2=r$. In (b) and (c) the fixation probability variation is given in terms of asymmetric differentiation rates $u_1=u_2=u$ and various values of $r$ and the ratio of the differentiation rates of normal SCs $\epsilon=u_2/u_1$. In (b) $\epsilon=0.5$ and in (c) $\epsilon=1.5$. }
\label{Noplast}
\end{figure}
%

\paragraph{Invasion in hierarchical model (zero plasticity).}
We now consider a more general case with no plastic potential, $\eta_{1,2}=0$. For the moment, we assume $r_1=\tilde{r}_1=1$, $r_2=\tilde{r}_2=r$.
\begin{eqnarray}
\rho_{\rm S}= \displaystyle{ \frac{1-  \frac{1-u_1}{r(1-u_2)} }{   1- \left(\frac{1-u_1}{r(1-u_2)}  \right)^{N_{\rm S}}}  },
\hspace{0.5cm} \rho_{\rm D}= 0, \hspace{0.5cm} \rho=\displaystyle{ \frac{ N_{\rm S}}{N_{\rm tot}} }\,\rho_{\rm S}. \label{Heir}
\end{eqnarray}
Fig. \ref{Noplast} shows how the survival chance of an initial mutant SC varies in terms of the relative fitness of mutant SCs/DCs as well as the probability of asymmetric division in the SC compartment. The population size is set to $N_{\rm S} = N_{\rm D} = 10$. The results are plotted for three sets of differentiation rates $(u_{1}=u_{2}=0.5),(u_{1}=0.1,u_{2}=0.5)$ and $(u_{1}=0.5,u_{2}=0.1)$ as $r$ varies (Fig. \ref{Noplast}(a)). If the normal cell differentiation rate $u_{1}$ is decreased from the balanced limit of $u_{1}=u_{2}=1/2$ the fixation probability decreases as well. Conversely, if the mutant cell differentiation rate, $u_{2}$ is decreased, away from $u_{2}=1/2$, $\rho_{\rm S}$ would increase.

We rewrite differentiation rates as $u_{1} = u$ and $u_{2} = \epsilon u$ where $u$ stands for an overall measure of differentiation, in both genotypes, and
$\epsilon$ measures the asymmetry in them. If $\epsilon <1$ the normal type differentiates more often per division and $\epsilon >1$ indicates that invader cells differentiate more often. Fig. \ref{Noplast}(b) shows $\rho_{\rm S}$ for $\epsilon =1/2$ as a function of $u$ for several division rates, $r=1,1.2,1.5,2$. For high differentiation rates, $u$, the fixation probability converges to unity. This is consistent with Eq. (\ref{Heir}). The relative fitness of the mutant versus normal types are given by $r(1-u_{2})/(1-u_{1}) = r(2-u)/(2(1-u))$. As $u \to 1$ the relative fitness approaches infinity and thus the value of fixation probability tend to unity.

Similar results are plotted for $\epsilon = 3/2$ in Fig. \ref{Noplast}(c). Now $\rho$ approaches zero for large $u$. Again this can be seen since $r(1-u_{2})/(1-u_{1}) = r(2-3u)/(2(1-u))$ which approaches zero as $u$ goes to 2/3.

\begin{figure}[h!]
 \centering
\subfigure[ ]{\includegraphics[width=0.49\textwidth]{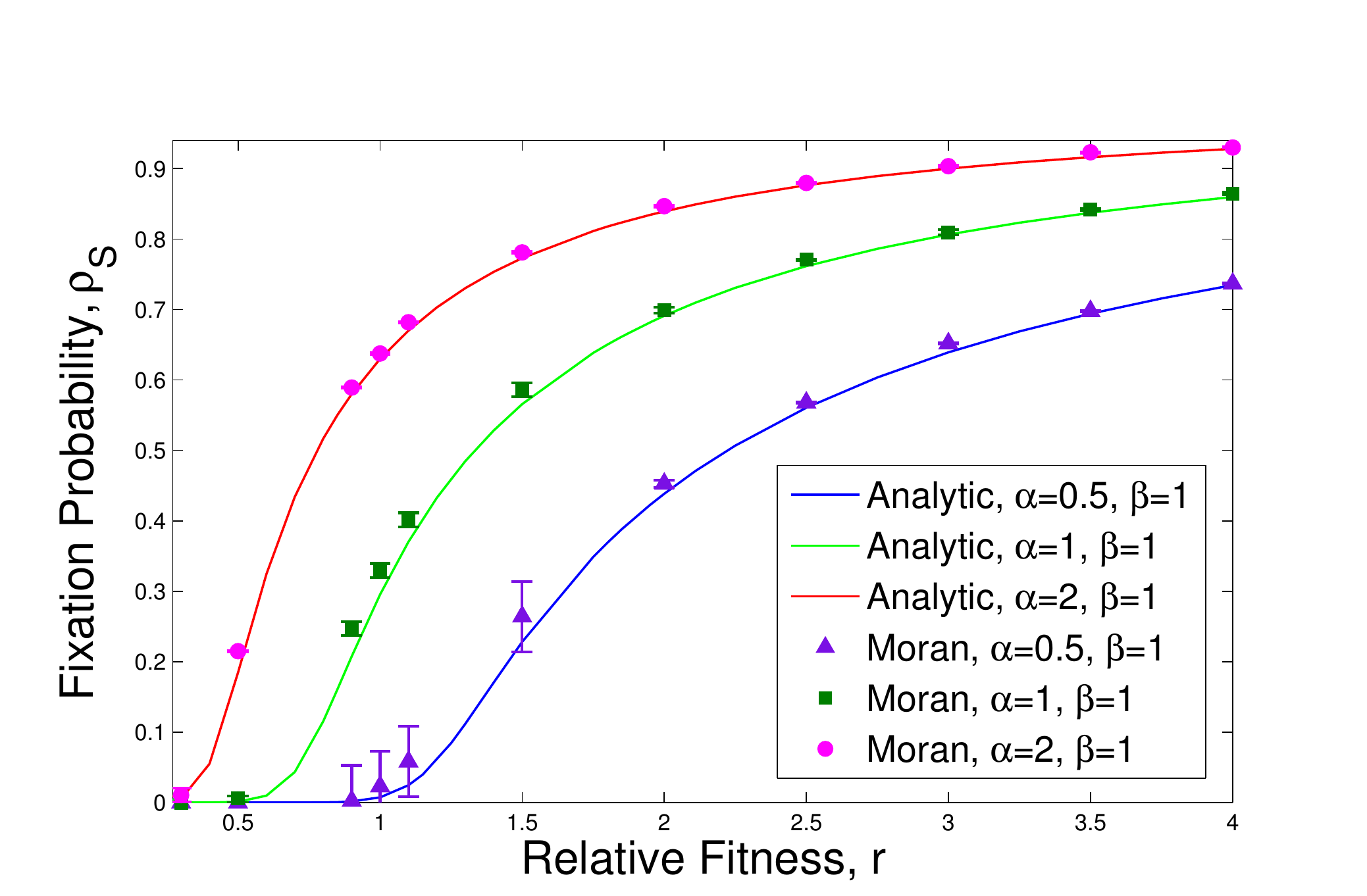} }
\subfigure[ ]{\includegraphics[width=0.47\textwidth]{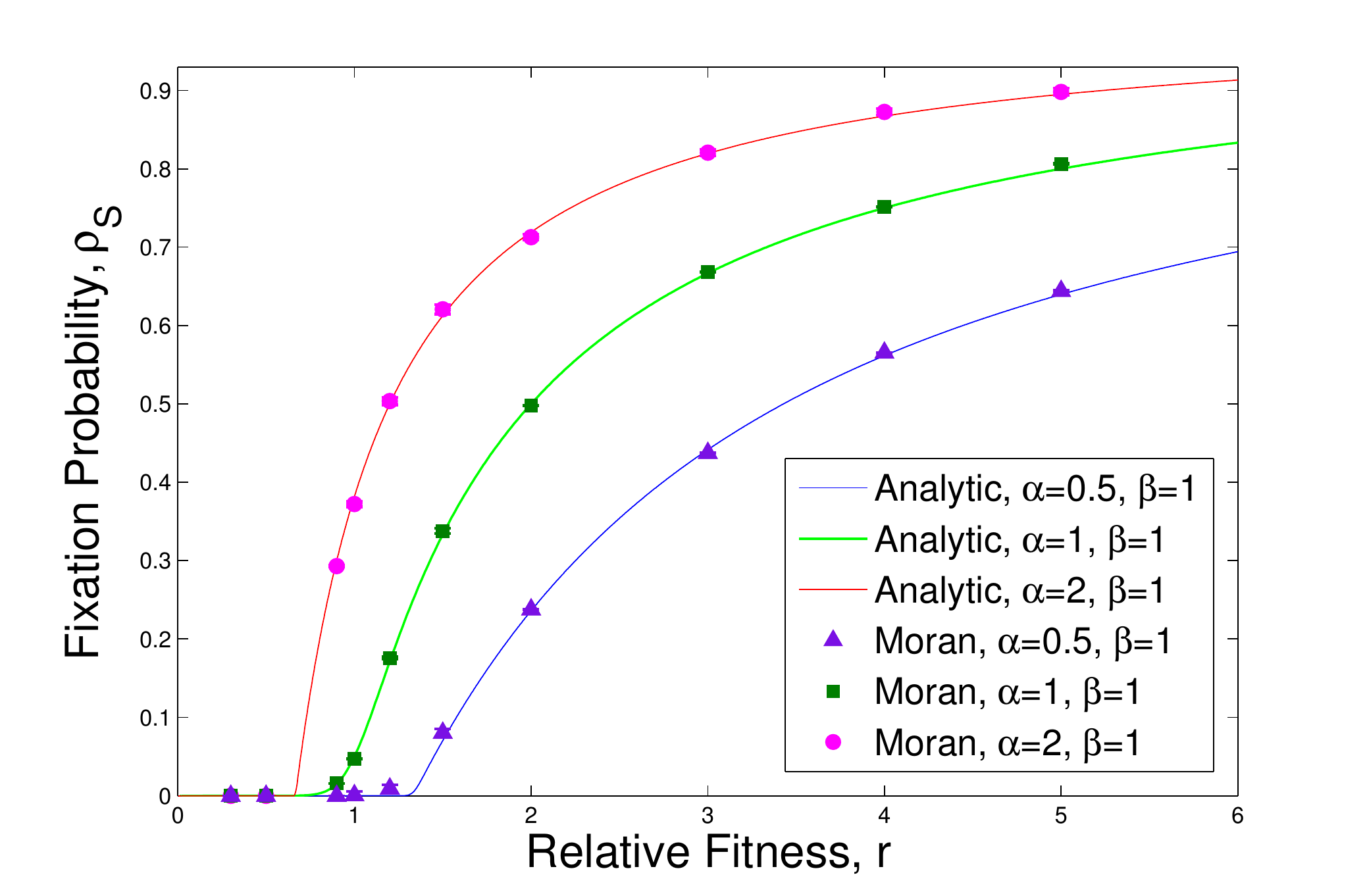} }
\caption {{\bf The fixation probability of mutants in the presence of phenotypic plasticity.} We suppose that $N_{\rm S}= N_{\rm D}=10$, $r_{1}= \tilde{r}_{1}= 1$, $r_{2}=\alpha r$, and $\tilde{r}_{2}=\beta r$ in the analytic results (solid curves) and stochastic simulation (points with bars as the standard deviation of mean). Changing $\alpha$ from 0.5 to 1 and then to 2 while $\beta$ remains fixed, the behavior of the system is shown in terms of the fixation probability with respect to the relative fitness $r$. In subfigure (a) plastic potential has only been considered for malignant individuals: $\eta_1=0, \eta_2=0.5$ while in (b) both WT and mutant cells can  dedifferentiate to the stemness state ($\eta_1=\eta_2=0.5$). Straightforward calculations reveal that for the given parameter values in (b), $\rho=\rho_{\rm S}=\rho_{\rm D}$.}
\label{figEtaU}
\end{figure}
%
\paragraph{Invasion in the presence of plasticity (dedifferentiation).}

Now we consider a more general case with non-zero differentiation and dedifferentiation rates. As before we set $r_{1}=\tilde{r}_{1}=1$. We parameterize the fitness of mutant subtypes as $r_{2}=\alpha r$ and $\tilde{r}_2=\beta r$. ($\alpha$ is different from the one introduced in the previous figure.) The parameter $r$ denotes the overall proliferation advantage of mutants over normal cells.

Fig. \ref{figEtaU} shows $\rho$ as a function of $r$ for various values of $\alpha$ and $\beta$. Fig. \ref{figEtaU}(a) assumes plastic mutants only ($\eta_{1}=0,\eta=0.5$) whereas Fig. \ref{figEtaU}(b) assumes equally plastic genotypes ($\eta_{1}=\eta_{2}$). The differentiation rates are equal and set to 1/2 for simplicity. As can be seen, the overall increase in $r$ increases the fixation probability monotonically in both cases. However, varying values of $\alpha$ and $\beta$ leads to the neutral value for the proliferation potential $r$. For example, in the case of Fig. (a) for $\alpha = \beta=1$, i.e. $r_{2}=\tilde{r}_{2}=r$, the mutant is advantaged for $r > 0.75$. For  $\alpha =0.5, \beta=1$, however, $r > 1.25$ has $\rho > 1/N$. Interestingly if stem cells divide faster than progenitors, i.e. $\alpha > \beta$, even for $r \approx 1/2$ the fixation probability is larger than the neutral value. We used the neutral fixation probability as $1/N_{\rm S}=1/N_{\rm D} =1/10$. Which is the neutral $\rho$ in the absence of (de)differentiation. Similar observations can be made in the case of $\eta_{1}=\eta_{2}$ in part (b). As can be seen our results are in very good agreement with exact stochastic simulations.

\begin{figure}[h!]
 \centering
\subfigure[] {\includegraphics[width=0.49\textwidth]{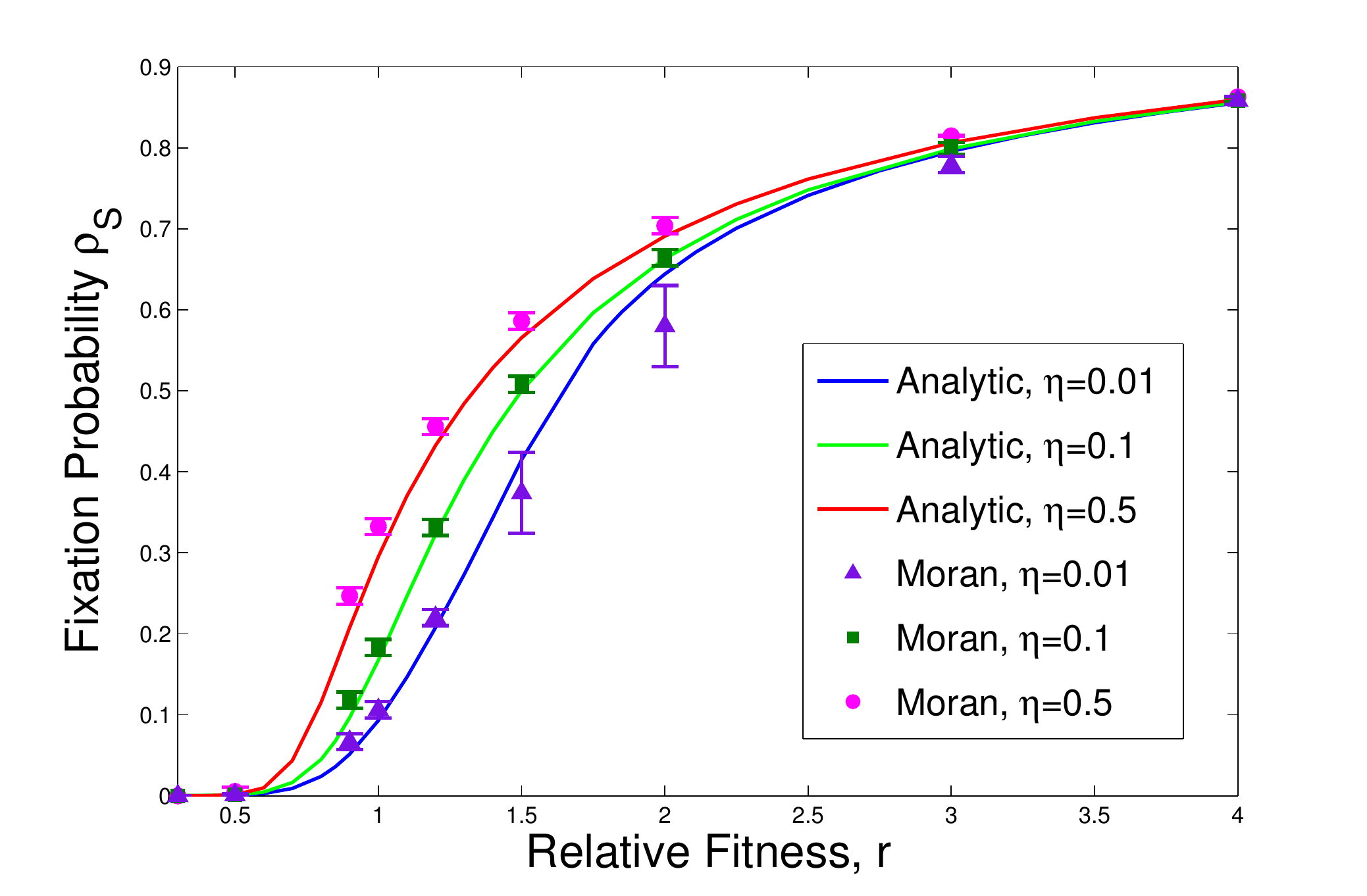} }
\subfigure[] {\includegraphics[width=0.49\textwidth]{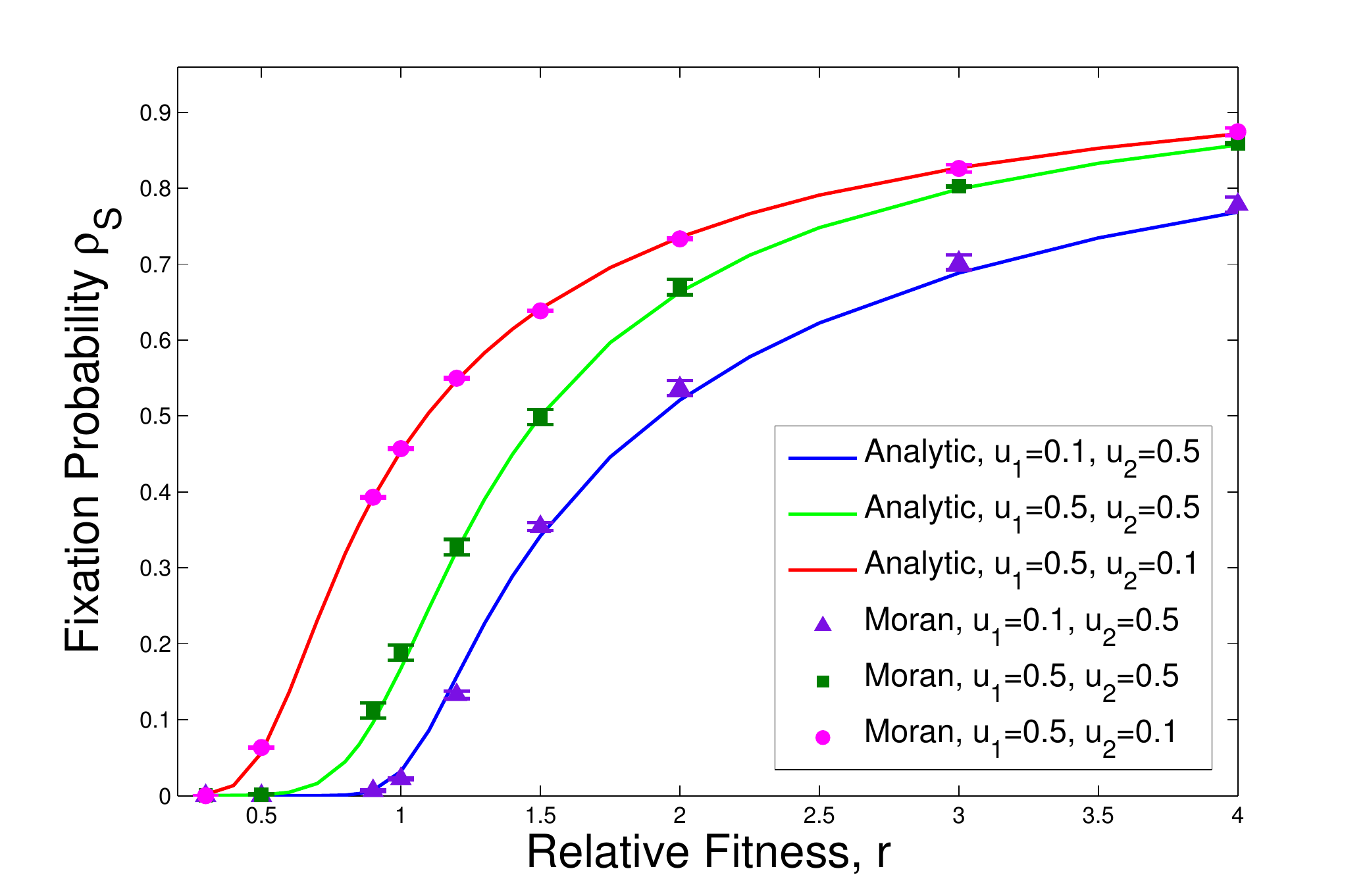}}
\caption { {\bf Effect of change in asymmetric differentiation and plasticity rates on survivability of mutants.} We assume that $N_{\rm S}= N_{\rm D}=10$, $r_1=\tilde{r}_1=1,$ and $r_2=\tilde{r}_2= r$. In subfigure (a), the fixation probability of SCs as a function of $\eta$ is given, where $\eta=0.01, 0.1, 0.5$ while $u_1=u_2=0.5,$ and $\eta_1=0$. In (b) $\eta_1=0$ and $\eta=0.1$. Changing parameters $u_1$ and $u_2$, which are the asymmetric division rates of normal and tumor SCs resp., the fixation probability as a function of $u_1, u_2$ is shown. Solid lines represent the analytic calculation and points correspond to simulation results (error bars are based on the standard deviation of mean).}
\label{ReplDyn}
\end{figure}

In Fig. \ref{figEtaU}, $\rho_{\rm S}$ is plotted as a function of $r$ for several values of $u_1$ and $u_2$ (part (a)) as well as $\eta$ (part (b)). The effect of dedifferentiation on increasing the value of the fixation probability is most significant for $r$ values corresponding to the neutral limit (Fig. \ref{figEtaU}(b)) and Fig. \ref{Noplast}(a). For example in Fig. \ref{figEtaU}(b) value of $\rho$ near $r=1$ increases from approximately $0.1 (\approx 1/N$) to 0.3 as $\eta$ increases to 0.5 (from 0.01). However for $r=4$ the difference between the two values of $\rho$ for $\eta=0.01, 0.5$ is negligibly small.

\begin{figure}[h!]
 \centering
\subfigure[ ]{\includegraphics[width=0.47\textwidth]{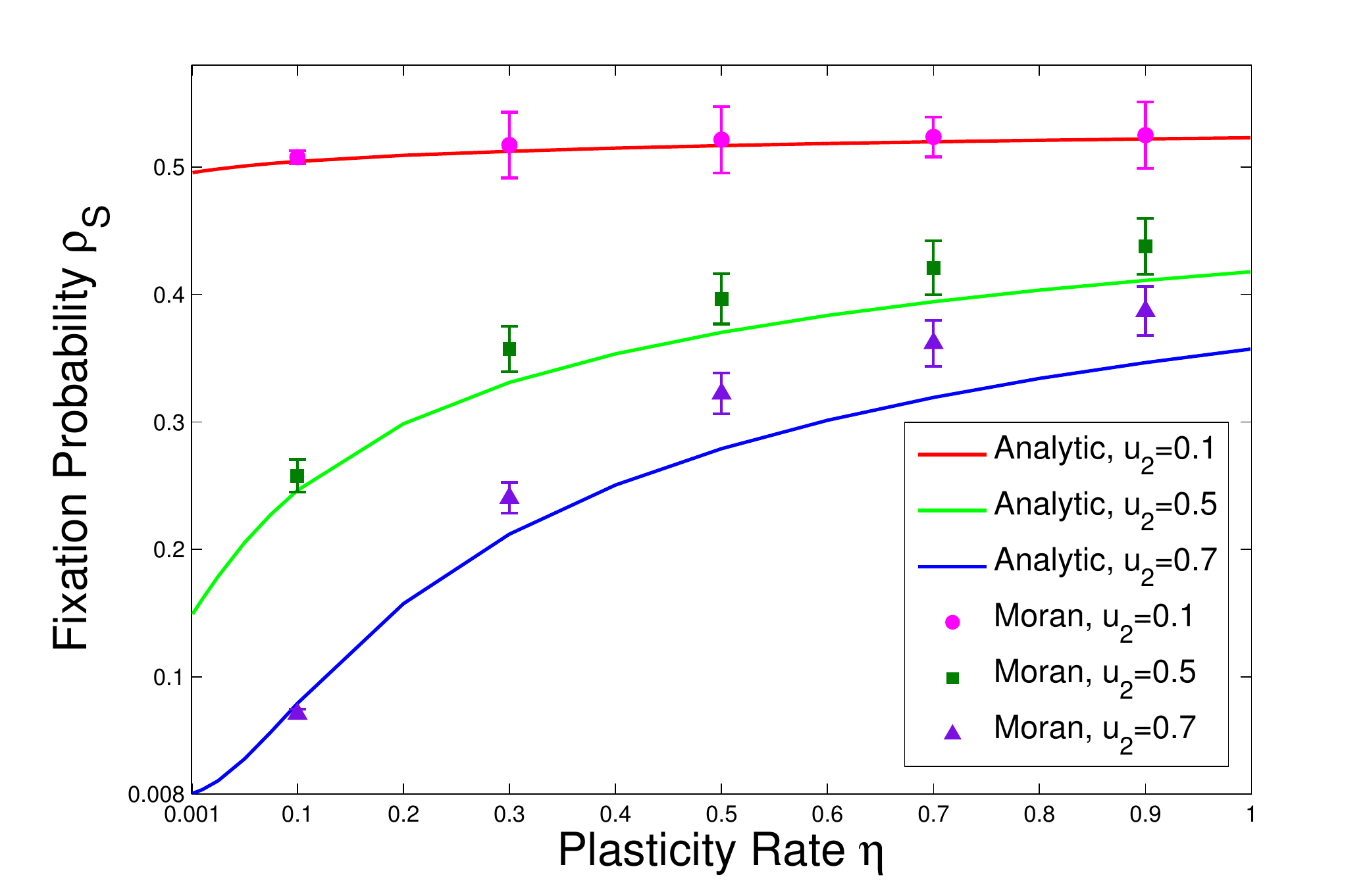} }
\subfigure[ ]{\includegraphics[width=0.47\textwidth]{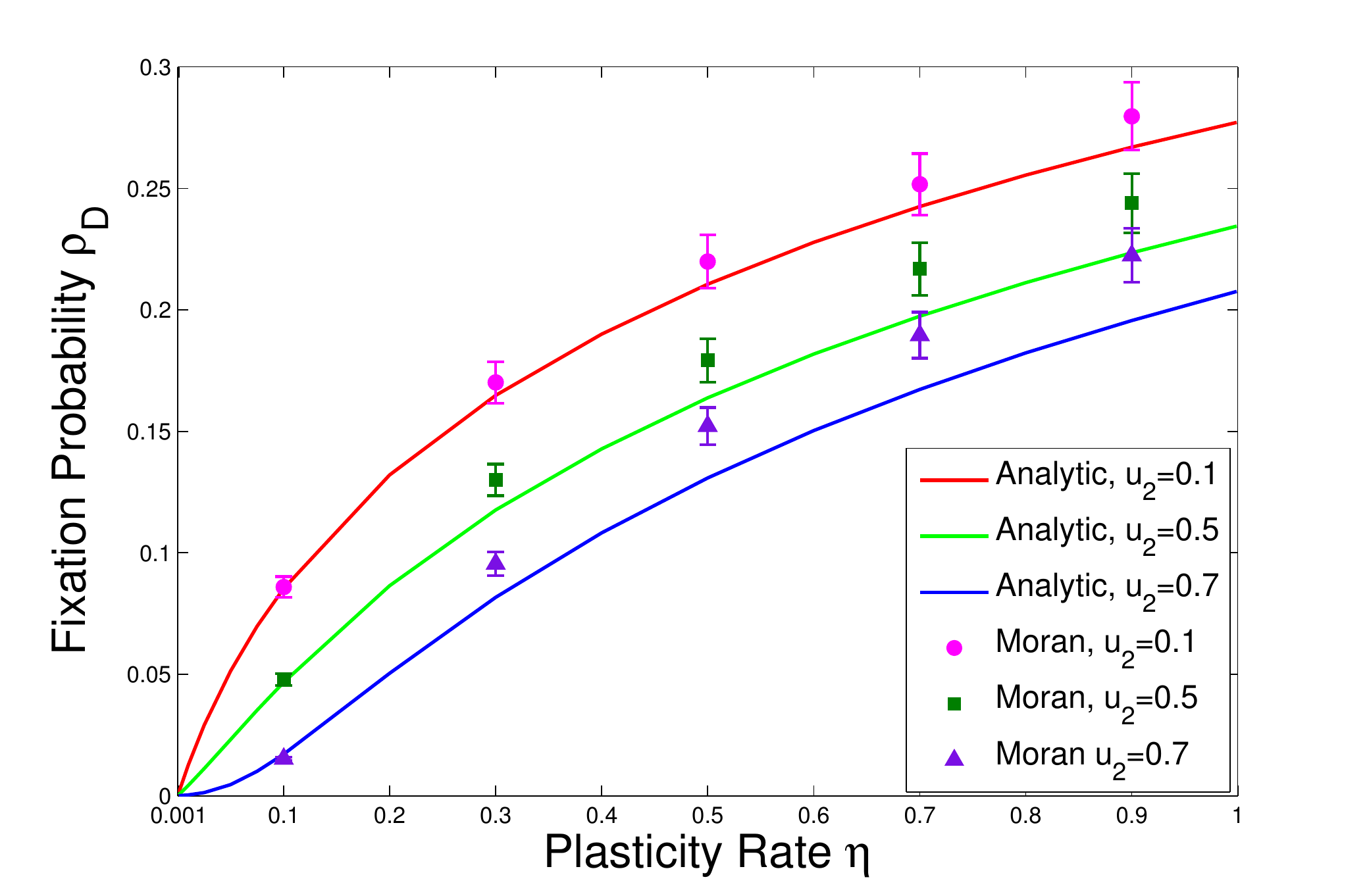} }\\[-2mm]
\subfigure[ ]{\includegraphics[width=0.47\textwidth]{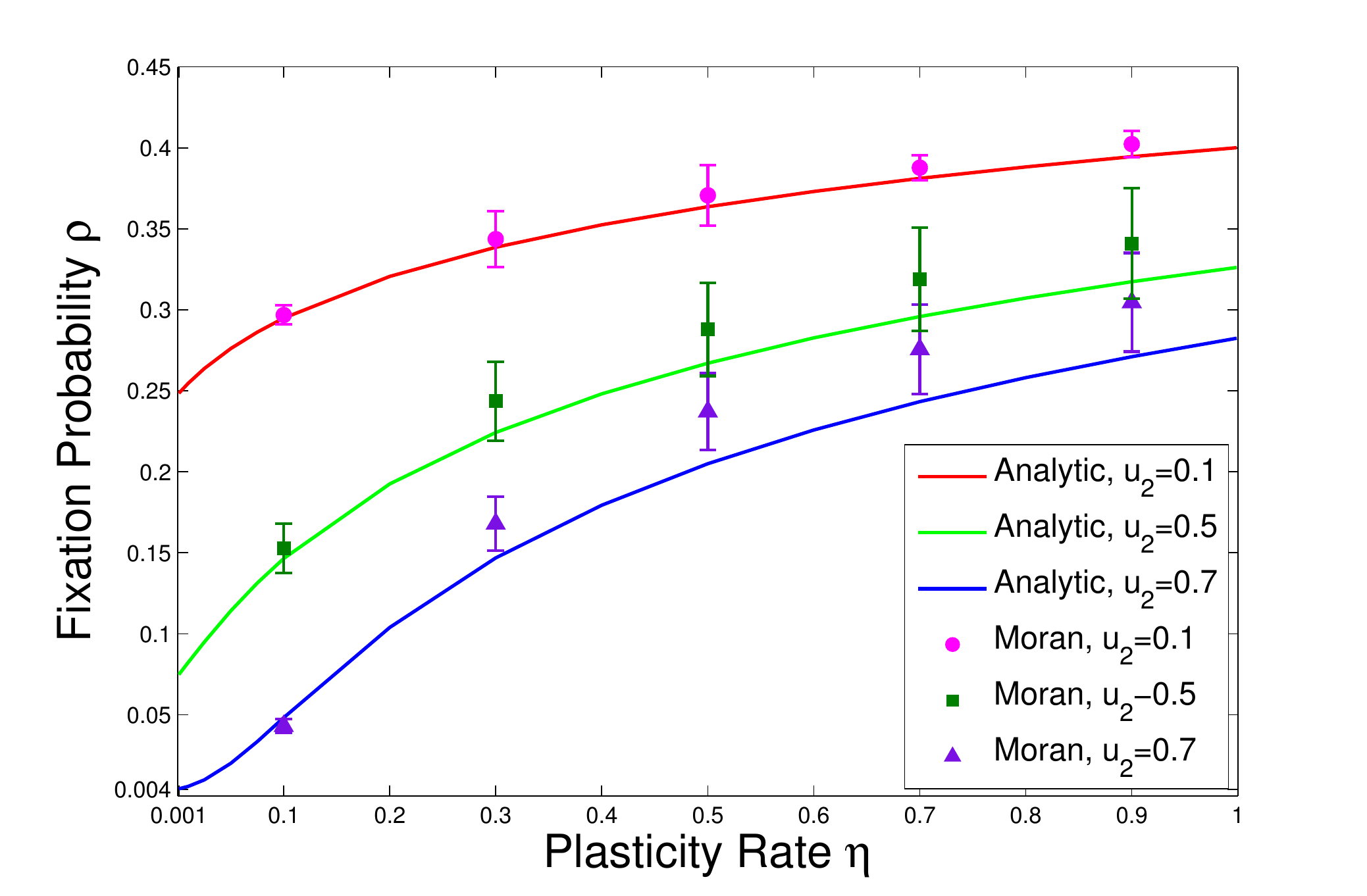} }
\caption { {\bf Effect of change in the rate of phenotypic plasticity for mutant DCs on the survival probability of malignant cells.} In subfigures (a), (b), and (c) we have resp. considered the fixation probability of mutants while the initial malignant mutation resp. occurs in SC, DC, and SC+DC compartments at random. In these plots, we assumed that $N_S=N_D=10$, $r_1=\tilde{r}_1=1$, $r_2=\tilde{r}_2=1.1, u_1=0.5$ and $\eta_1=0$. Different asymmetric division rates of normal and malignant individuals have also taken into account when the fixation probability is given as a function of  $\eta$. Solid curves and points are respectively the results of analytic and simulation analysis. At each given discrete point, the error bar depicts the standard deviation of mean.}
\label{FPeta}
\end{figure}

In the presence of dedifferentiation, mutations arising in the differentiated compartment can now undergo clonal expansion. In Fig. \ref{FPeta} we plotted values of the fixation probabilities $\rho_{\rm S}$, $\rho_{\rm D}$ and the average fixation $\rho$ as a function of $\eta$. As expected for $\eta=0$, $\rho_{\rm D}$ tends to zero for various values of $u_{1}$ and $u_{2}$. However $\rho_{\rm S}$ approaches the results obtained before (Eq. (\ref{Heir})). In Fig.\ref{FPeta}(a) for $u_1=u_2=0.5$ and $r=1.1$ $\rho_{\rm S}$ approaches 0.15 as $\eta=0$. This is in agreement with the extended Moran result, Eq. (\ref{Heir}), with effective fitness $r(1-u_2)/(1-u_1)$ which for $N_{\rm S}=10$ gives $\rho_{\rm S}=0.15$. As $\eta$ increases $\rho_{\rm S}$ increases further and approaches 0.4 for large values of $\eta$. We can see that finite values of $\eta$ have now conferred a selection advantage on the previously deleterious mutants. For example for $u_{1}=0.5,u_{2}=0.1$ and $r=1.1$, the effective fitness is less than unity with $\rho_{\rm S}  \approx 0$. For finite values of the dedifferentiation rate, $\rho_{\rm S}$ will exceed the neutral value $1/N$. For example for $\eta=0.5$ we can read $\rho_{\rm S} \approx 0.25$ in this case.

\subsection*{\bf Phase diagram}

To derive the phase diagram that determines the evolutionary advantage of a malignant genotype, we begin with the system of equations (\ref{eq:RepM4}). The fixed points of the replicator equation determine the stationary frequencies of mutant subtypes: $x^{\star}_{\rm S},x^{\star}_{\rm D}$. There are four fixed--points for the system of equations (\ref{eq:RepM4}). The fixed point $(x^{\star}_{\rm S}=0,x^{\star}_{\rm D}=0)$ corresponds to the extinction of both phenotypes and other fixed points $(x^{\star}_{\rm S}=1,x^{\star}_{\rm D}=1)$ or any $(x^{\star}_{\rm S}\neq 0,x^{\star}_{\rm D} \neq 0)$ corresponds to successful invasion by the mutant. We do not distinguish between these fixed points as the stochastic fluctuations prohibit coexistence of mutant and WT phenotypes. We determine the phase boundary when a fixed point $(x^{\star}_{\rm S},x^{\star}_{\rm D})$ merges with $(0,0)$. This indicates that the invasion has become unstable to parameter changes, and determines the phase boundary in the phase space of parameters.

Similar results can be obtained from the Eqs. (\ref{FP})--(\ref{FP2}) requiring that $\rho_{\rm S}=1/N_{\rm S}$ (corresponding to the neutral selection). In a large population, this approach would be identical to the results from the replicator equation.

At first, we assume $u_1=u_2=\eta_1=\eta_2 =0$, then the solution to the system (\ref{eqRD}) results in two fixed points, $(x^{\star}_{\rm S}=0,x^{\star}_{\rm D}=0)$ (extinction) or $(x^{\star}_{\rm S}=1,x^{\star}_{\rm D}=1)$ (fixation) of invading mutant SCs (or DCs).

%

\begin{figure}[h!]
 \centering
\includegraphics[angle=270,width=0.95\textwidth]{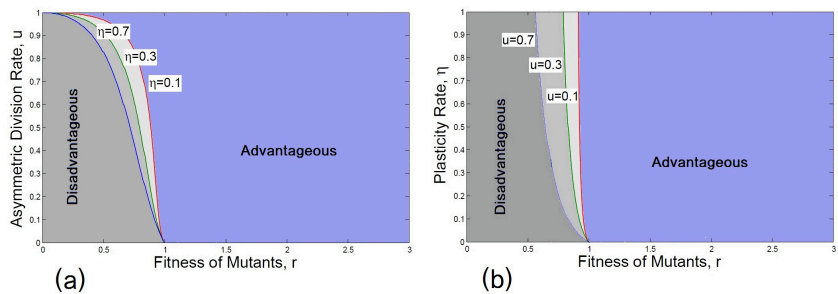}\\[-2cm] 
\caption { {\bf Phase diagram of plastic mutant SCs.} The phase boundary for advantageous and disadvantageous mutant populations are given as differentiation and plasticity rates change. We assume that $r_1=\tilde{r}_1=1$, $r_2=\tilde{r}_2=r$, $u_1=u_2=u$, and $\eta_1=0$. Different regions for advantageous and disadvantageous mutant SCs are given in (a) as $u$ changes. A similar analysis has been carried out in (b) as $\eta$ varies. In (a) $\eta=0.1,0.3,0.7$, here the alteration in the plasticity rate of DCs results in a tendency to approach various regions of fixation for mutant SCs, while the extinction domain shrinks with increasing $\eta$. In (b) $u=0.1, 0.3, 0.7$. Increasing the asymmetric division rate $u$, the region for advantageous mutants expands to provide a higher survival chance for mutant SCs. In both cases, advantageous criteria relate to either fixation of mutants or coexistence of mutants and WT individuals.}
\label{phase}
\end{figure}
%
Next, we consider the case with $r_1= \tilde{r}_1= 1,$  $r_2= \tilde{r}_2= r$, $u_1= u_2= u,$  $\eta_1= 0,$ and $\eta_2=\eta$. We have plotted the phase diagram for the model parameter space of $\eta-u-r$. The phase boundary that determines the condition for neutrality is given by the algebraic relation (see SI for details)
\begin{eqnarray}
(u+\eta-1)r^2+\left[-u^2+(\eta-1)u+2-\eta \right]r-1+u^2=0,
\end{eqnarray}

The phase diagram is plotted in the space of $u-r$ (Fig. \ref{phase}(a)) and $\eta-r$ (Fig.\ref{phase}(b)). The observation that an increase in $\eta$ can turn a previously deleterious mutant into a beneficial one, can be seen here as well. For example in Fig.\ref{phase}(b) for $u=0.7$ and $r = 0.9 < 1$ where we expect a deleterious mutant for large enough values of $\eta$, we can cross the phase boundary into an advantaged region. Interestingly we observe a similar trend as the overall differentiation rate increases.

\subsection*{\bf Application to colorectal cancer}

Clonal expansion in colorectal cancer is known to be initiated as a result of mutations occurring at the bottom of the crypt \citep{vogelstein2013cancer}.  More recently, Vermeulen {\it et al} \citep{Ver} considered the dynamics of cells at the bottom of a normal colonic/intestinal crypt. Such controlled cells in their circular model of size 5 undergo a selection process. The selection mechanism is investigated for several oncogenes and tumor suppressor genes imposed at the crypt base. Fixation probability of a single mutation and the relative fitness $r$ of the mutant cells, compared with the normal host, are reported. Their estimates reveal that the relative fitness of the original cell containing APC$^{-/-}$ is $r=1.58$, while $r=1.56$ for Kras$^{G12D}$, and $r=0.96$ for P53$^{R172H}$ (compared with normal control cells mice). P53 mutation seems not to confer a fitness advantage and is weakly deleterious. However for P53$^{R172H}$, it has been observed that the fitness of a mutant elevates from $0.96$ to $1.16$ in comparison with the DSS-treated cells (colitis) which is in the presence of inflammatory injury. Thus, under inflammatory signaling effects, the P53$^{R172H}$ mutants appear to gain a selection advantage and thus a higher fitness \citep{Ver,VerN}.

According to the recent {\it in vivo} study by Schwitalla {\it et al} \citep{schwitalla2013intestinal}, inflammatory signaling plays a role in elevating the rate of dedifferentiation. It has been also shown that inflammatory disease activates the transcription factor NF-${\kappa}$B. NF-${\kappa}$B can, in turn, elevate Wnt-signaling which leads to the phenotypic plasticity of non-SCs \citep{karin2005nf}.

Thus we suggest that the higher survival chance of the P53 mutated SCs along with the DSS-treated cells, may be the result of dedifferentiation in the presence of inflammatory stroma \citep{schwitalla2013intestinal}. The survival probability of mutants in colon/intestine, in the presence of inflammatory signaling, is presumably correlated with both the fitness and plastic nature of the epithelial cells. Thus the fixation probability reported in \citep{Ver} of cells with P53 DSS colitis mutation, can be derived by the same fitness $r=0.96$ for P53 when dedifferentiation occurs.

\begin{figure}[H]
\centering
\includegraphics[angle=270,width=0.6\textwidth]{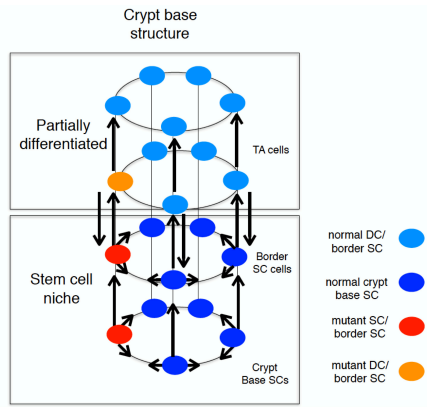}\\[2mm]
\caption {{\bf Cellular interactions in the colonic crypt as a newborn mutant arises within the Stem or differentiated compartments.} Within this schematic cylindrical model, we represent how our model is structured through the four compartments of host and mutant stem and differentiated cells. In contrast to the circular model of five SCs considered in \citep{VerN}, we assume a cylindrical model of two circles, one on the top of the other. SCs are located at the bottom circle while the circle on the top is full of partially-differentiated cells.}
\label{crypt}
\end{figure}

For instance, when $r=0.96$ and $u_1=0.5, u_2=0.25$, having the plasticity rate at $\eta= 0.12$, one obtains the same fixation probability as \citep{Ver} (see Table \ref{table1}). This finding strongly suggests that there is an elevation in the fixation of a deleterious mutant into an advantageous trait due to the plastic properties of the mutant (see also the next section). In Fig. \ref{crypt}, we considered a cylindrical model of the crypt-base, and immediate adjoining transit amplifier layers, cells in the colon/intestine. The bottom layer consists of central stem cells and border stem cells, while transit ampflifying and non-SC cells reside on top of this layer and at higher layers. A mouse crypt is comprised of 5--7 functional stem cells \citep{potten1992proliferation,Ver}. To compare the results of our model with experimental observations of \citep{Ver} for the fixation probability of P53 DSS colitis, we consider the first two circular layers of SCs (see Fig. \ref{crypt}) in which we roughly assume $N_{\rm S} \approx 10$  \citep{RL} and note that the fitness reported in the experimental data \citep{Ver} is for functional SCs and may be considered unchanged for the two layers ($N_{\rm S}=10$) of central (functional) and border SCs as well.
\begin{table}[h!]
    \centering
    \caption{{\bf Comparison between our analytic result and experiment \citep{Ver}} 
    }
\label{table1}
    \begin{tabular}{l l l l  l}
        \hline
        \rowcolor[rgb]{0.8,0.8,1}&$r$&$\eta_2$&$ \rho_{\rm S}$ & Reference\\\hline
      P53 Controlled & 0.96 &   0    & 0.113 & \citep{Ver} \\\hline
      P53 DSS Colitis & 1.16 &   0    & 0.343& \citep{Ver} \\\hline
      P53 DSS Colitis+Plasticity   & 0.96 &  0.12   & 0.343& Results Section\\\hline
\end{tabular}

\end{table}
%
\section*{Discussion}

The evolutionary implications of epigenetic heterogeneity are not very well understood in cancer biology. A known picture for phenotypic heterogeneity (when the genotypes are assumed to be identical) relates to the cancer stem cell hierarchy. In this picture,  pluripotent cells with tumor initiating capacity can undergo mitotic events and either replenish their own population or produce a lineage of partially differentiated cells (including precursor and/or transit amplifying cells).

In the current study, we present a general model of four distinct subpopulations to investigate Darwinian evolution in such a hierarchical structure. We consider two genotypically different populations (mutant and wild-type). Mutations are results of unwanted oncogenic or TSG mutations. Each genotype has phenotypically different subtypes of stem cells (SCs) and differentiated cells (DCs). SCs ultimately generate their associated progenies, DCs, through proliferation and asymmetric differentiation. DCs have restricted  proliferation capacity. Due to the tissue structure of the crypt, the population of different subtypes remains approximately unchanged. Genetic mutations can occur among SCs or DCs which we assume to be occurring through a uniform probability. Mutations can confer not only higher division rates but also different rates of differentiation and plasticity among mutant subtypes. These changes can be triggered, for example, by microenvironmental conditions.

Our model predicts the fixation probability of a newborn mutant as a function of  the division rates of mutant and resident SCs/DCs, differentiation and dedifferentiation rates. Exact analytic calculation and numerical simulations -- which are in almost perfect agreement -- suggest that the asymmetric differentiation in the SC group has a major effect on the fate of mutants compared with dedifferentiation. More specifically, a greater impact on the fixation probability of SCs can be observed by the change in asymmetric differentiation of normal cells compared with that of malignant cells. Furthermore, we observe that the more plastic trait has an evolutionary advantage. This is most notable close to the neutral limit, i.e. when the proliferation rates of the mutants and residents are very close in value. Most interestingly, a sufficient increase in the rate of plasticity can turn a previously deleterious mutant into a beneficial one.

As an important application of this model, we consider the intestinal/colonic crypt with two groups of SCs and their neighboring, partially differentiated cells. The competition between malignant mutations and normal cells in the base of the crypt has attracted much research interest, and is one the most studied scenarios in cancer evolution. It has recently been shown that the Moran type process, initially suggested by \citep{Kami,kaveh2015duality,Kam,Kam1,N,N1} are in perfect agreement with the experimental observation \citep{VerN}. In this {\it in vivo} experimental analysis, ${\rm KRAS}$, ${\rm APC}^{+/-}$, ${\rm APC}^{-/-}$, and ${\rm P53}$ DSS colitis mutations separately induced in the crypt base. Assessing clonal lineage tracing \citep{VerN}, the authors were able to observe the growth of mutants in the populations and thus measured the fixation probability of mutants. However, the experimental observations sometimes ignore the microenvironmental interaction between the neighboring transit amplifying (TA) cells and SCs in the crypt (See \citep{RL,VerN} for the models and estimated parameter values of population dynamics of the crypt, as well as numbers of central and border SCs at the base of the crypt).

In this investigation, we have suggested a plausible estimation method for the dedifferentiation rate which can give the same value of the fixation probabilities of P53 inactivation in the crypt base as observed by \citep{Ver}. As the authors in \citep{Ver} noticed, P53 mutations can maintain advantageous features to invade the crypt base in the presence of inflammatory signaling. We propose that this could potentially be a result of dedifferentiation caused by an inflammatory stroma \citep{schwitalla2013intestinal}. This finding supports our idea about the elevation of a deleterious mutant into one with an advantageous trait due to the plastic properties of the mutant.

This research is one of many mathematical approaches recently used to investigate various aspects of tumorigenesis. In our approach, we attempt to dissect a part of the complex machinery in multi--compartmental models, and understand this in greater detail.
This study provides various insights concerning some features of initiation and progression of cancer, and suggests possible experimental investigations to confirm some of the theoretical results. The approach taken in this paper may lead to a better understanding of the natural pathological mechanisms in the colonic/intestinal epithelium or any similar four--compartmental structures in ecology, population genetics, and social networks. The particular application of this approach to carcinogenesis may have some results at prognosis.

\section*{Acknowledgments}

This work was financially supported by the Natural Sciences and Engineering Research Council of Canada (NSERC, discovery grant to SS and MK).



\newpage

\section*{Supporting Information}

\subsection*{A. Characteristic equation}
Starting from the master equation of the given model (where for $N=N_{\rm S}+N_{\rm D}$, we assume that $1/N$ is the duration of each updating time)
\begin{eqnarray}
\frac{\partial p(n_{\rm S},n_{\rm D};t)}{N\,\partial t} \!\!&=&\!\!  W^{+}_{{\rm S}}(n_{\rm S}-1,n_{\rm D})\, p(n_{\rm S}-1,n_{\rm D};t) +   W^{-}_{{\rm S}}(n_{\rm S}+1,n_{\rm D})\,p(n_{\rm S}+1,n_{\rm D};t)\nonumber\\
&& \hspace{-2cm} +\, W^{+}_{{\rm D}}(n_{\rm S},n_{\rm D}-1)\, p(n_{\rm S},n_{\rm D}-1;t) +   W^{-}_{{\rm D}}(n_{\rm S},n_{\rm D}+1)\, p(n_{\rm S},n_{\rm D}+1;t)  \nonumber \\[2mm]
&& \hspace{-2cm} -\, \big( W^{+}_{{\rm S}}(n_{\rm S},n_{\rm D})+W^{+}_{{\rm D}}(n_{\rm S},n_{\rm D})+W^{-}_{{\rm S}}(n_{\rm S},n_{\rm D})+W^{-}_{{\rm D}}(n_{\rm S},n_{\rm D})\big)\, p(n_{\rm S},n_{\rm D};t),\label{eq:master11}
\end{eqnarray}
the generating equation can be derived by assuming the probability generating function (PGF) in which coefficients define the probabilities of being at different possible states after a given time.

Starting from the master Eq. (\ref{eq:master11}) discussed in the previous section, we can analyze the probability of absorption (fixation or extinction) using generating function techniques recently developed for the constant population birth-death processes \citep{HV1}. Defining the probability generating function (PGF) of the probability density function $p(n_{\rm S},n_{\rm D} ; t)$ of having $n_{\rm S}$ mutant SCs and $n_{\rm D}$ mutant DCs at time $t$. Then the PGF is
\begin{eqnarray}
F(z_{\rm S},z_{\rm D};t) = \sum_{n_{\rm S},n_{\rm D}}z^{n_{\rm S}}_{\rm S}z^{n_{\rm D}}_{\rm D}\, p(n_{\rm S},n_{\rm D};t)
\end{eqnarray}
Using the PGF one can rewrite the master equation (the latter equation) for PGF to derive the characteristic equations  as follows
\begin{eqnarray}
\frac{\partial F}{N\,\partial t} =  (z_{\rm S}^{-1}-1)\left\langle (W^-_{\rm S} - z_{\rm S}\,W^+_{\rm S})\,z_{\rm S}^{n_{\rm S} } z_{\rm D}^{n_{\rm D}}  \right\rangle + (z_{\rm D}^{-1}-1)\left\langle (W^-_{\rm D} - z_{\rm  D}\,W^+_{\rm D})\,z_{\rm S} ^{n_{\rm S} } z_{\rm D} ^{n_{\rm D} }  \right\rangle, \label{pgf1}
\end{eqnarray}
where the operators $W^{\pm}_{\rm S}$ and $W^{\pm}_{\rm D}$ are
\begin{eqnarray}
W^{+}_{{\rm S}}(n_{\rm S},n_{\rm D}) &=& {\rm Prob}(n_{\rm S},n_{\rm D} \rightarrow n_{\rm S}+1,n_{\rm D})\nonumber \\
&=& \left( \frac{r_{2}\,(1-u_{2})\,n_{\rm S} +  \tilde{r}_{2}\,\eta_{2}\, n_{\rm D}}{  r_{1}\,(N_{\rm S}-n_{\rm S})  + r_{2}\, n_{\rm S} + \tilde{r}_{1}(N_{\rm D}-n_{\rm D}) +  \tilde{r}_{2}\, n_{\rm D} } \right) \, \frac{N_{\rm S}-n_{\rm S}}{N_{\rm S}}, \nonumber\\
W^{-}_{{\rm S}}(n_{\rm S},n_{\rm D})  &=& {\rm Prob}(n_{\rm S},n_{\rm D} \rightarrow n_{\rm S}-1,n_{\rm D})\nonumber \\
&=& \left( \frac{r_{1}\, (1-u_{1})\, (N_{\rm S}-n_{\rm S})  + \tilde{r}_{1}\, \eta_{1}\, (N_{\rm D}-n_{\rm D})}{    r_{1}\,(N_{\rm S}-n_{\rm S})  + r_{2}\, n_{\rm S} + \tilde{r}_{1}(N_{\rm D}-n_{\rm D}) +  \tilde{r}_{2}\, n_{\rm D} } \right) \, \frac{n_{\rm S}}{N_{\rm S}}, \nonumber\\[-3mm]
&& \label{eq9} \\[-2mm]
W^{+}_{{\rm D}}(n_{\rm S},n_{\rm D})  &=& {\rm Prob}(n_{\rm S},n_{\rm D} \rightarrow n_{\rm S},n_{\rm D}+1)\nonumber \\
&=& \left( \frac{ \tilde{r}_{2}\, (1-\eta_{2})\, n_{\rm D}+  r_{2}\, u_{2}\, n_{\rm S}}{   r_{1}\,(N_{\rm S}-n_{\rm S})  + r_{2}\, n_{\rm S} + \tilde{r}_{1}(N_{\rm D}-n_{\rm D}) +  \tilde{r}_{2}\, n_{\rm D} } \right)\, \frac{N_{\rm D}-n_{\rm D}}{N_{\rm D}}, \nonumber\\
W^{-}_{{\rm D}} (n_{\rm S},n_{\rm D}) &=& {\rm Prob}(n_{\rm S},n_{\rm D} \rightarrow n_{\rm S},n_{\rm D} -1)\nonumber \\
&=&  \left( \frac{ \tilde{r}_{1}\, (1-\eta_{1})\, (N_{\rm D}- n_{\rm D})  +   r_{1}\, u_{1}\, (N_{\rm S}-n_{\rm S}) }{   r_{1}\,(N_{\rm S}-n_{\rm S})  + r_{2}\, n_{\rm S} + \tilde{r}_{1}(N_{\rm D}-n_{\rm D}) +  \tilde{r}_{2}\, n_{\rm D}   } \right)\, \frac{n_{\rm D}}{N_{\rm D}}. \nonumber
\end{eqnarray}
$W^{\pm}_{{\rm D}} (n_{\rm S},n_{\rm D})$ and $W^{\pm}_{{\rm D}} (n_{\rm S},n_{\rm D})$ define the transition probabilities for increase/decrease in the number of mutant SCs and DCs resp. Acquiring these probabilities with those for no change in the number of mutant cells in either compartments, one can define the transition matrix which incorporates random walks among various states prior to  fixation.

Substituting Eqs. (\ref{eq9}) into generating equation (\ref{pgf1}), we obtain:
\begin{eqnarray}
\frac{\partial F}{N\,\partial t}
&=& \Big\langle \frac{ (z^{-1}_{\rm S}-1)(r_{2}\,(1-u_{2})\,n_{\rm S} +  \tilde{r}_{2}\,\eta_{2}\, n_{\rm D}) (N_{\rm S}-n_{\rm S})}{ (r_{1}\,(N_{\rm S}-n_{\rm S})  + r_{2}\, n_{\rm S} + \tilde{r}_{1}(N_{\rm D}-n_{\rm D}) +  \tilde{r}_{2}\, n_{\rm D})\,N_{\rm S} }  \Big\rangle \nonumber\\[2mm]
&& +\, \Big\langle\frac{  (1-z_{\rm S})(r_{1}\, (1-u_{1})\, (N_{\rm S}-n_{\rm S})  + \tilde{r}_{1}\, \eta_{1}\, (N_{\rm D}-n_{\rm D}))\,n_{\rm S}  }{   N_S(r_{1}\,(N_{\rm S}-n_{\rm S})  + r_{2}\, n_{\rm S} + \tilde{r}_{1}(N_{\rm D}-n_{\rm D}) +  \tilde{r}_{2}\, n_{\rm D}) } \Big\rangle  \nonumber\\[2mm]
&&  +\, \Big\langle  \frac{ (z^{-1}_{\rm D}-1)(\tilde{r}_{2}\, (1-\eta_{2})\, n_{\rm D}+  r_{2}\, u_{2}\, n_{\rm S}) (N_{\rm D}-n_{\rm D})  }{  (r_{1}\,(N_{\rm S}-n_{\rm S})  + r_{2}\, n_{\rm S} + \tilde{r}_{1}(N_{\rm D}-n_{\rm D}) +  \tilde{r}_{2}\, n_{\rm D})\,N_{\rm D}  } \Big\rangle\nonumber\\[2mm]
&& +\, \Big\langle \frac{  (1-z_{\rm D})(\tilde{r}_{1}\, (1-\eta_{1})\, (N_{\rm D}- n_{\rm D})  +   r_{1}\, u_{1}\, (N_{\rm S}-n_{\rm S}))\,n_{\rm D}  }{ (r_{1}\,(N_{\rm S}-n_{\rm S})  + r_{2}\, n_{\rm S} + \tilde{r}_{1}(N_{\rm D}-n_{\rm D}) +  \tilde{r}_{2}\, n_{\rm D})\,N_{\rm D}  } \Big\rangle \nonumber\\
\end{eqnarray}
Defining $\hat{n}_{\rm S} = z_{\rm S}\frac{\partial}{\partial z_{\rm S}},    \hat{n}_{\rm D} =  z_{\rm D}\frac{\partial}{\partial z_{\rm D}}$, we conclude
\begin{eqnarray}
\frac{\partial F}{N\,\partial t} &=&   \left\{ (z^{-1}_{\rm S}-1)W^{-}_{\rm S}(\hat{n}_{\rm S},\hat{n}_{\rm D})  + (z_{\rm S}-1)\, W^{+}_{\rm S}(\hat{n}_{\rm S},\hat{n}_{\rm D}) \right. \nonumber\\[2mm]
&& \left. + (z^{-1}_{\rm D}-1)W^{-}_{\rm D}(\hat{n}_{\rm S},\hat{n}_{\rm D}) +  (z_{\rm P}-1)W^{+}_{\rm D}(\hat{n}_{\rm S},\hat{n}_{\rm D})\,\right\}\,F.
\end{eqnarray}
For large -- $N_{\rm S}$ ($N_{\rm D}$), we can simplify the latter equation for the probability generating function by keeping the linear derivative terms to leading order in $N_{\rm S}$ ($N_{\rm D}$). This tends to
\begin{eqnarray}
\frac{\partial F}{N\,\partial t} &\simeq& (z_{\rm S}-1)\left\{   \frac{  r_{2}(1-u_2) \, z_{\rm S}\, \frac{ \partial F} { \partial z_ {\rm S}} + \tilde{r}_2\, \eta_2\, z_{\rm S}\, \frac{ \partial F} { \partial z_ {\rm D}}  }{\hat{N}_{\rm r, S, D}}    - \frac{  r_1\,(1-u_1)\, N_{\rm S}\, \frac{ \partial F} { \partial z_ {\rm S}} + \tilde{r}_1\, \eta_1\, N_{\rm D}\, \frac{ \partial F} { \partial z_ {\rm S}}  }  { \hat{N}_{\rm r, S, D}\, N_{\rm S} }   \right\}   \nonumber\\
&& \hspace{-1.5cm} + (z_{\rm D}-1)\left\{   \frac{  \tilde{r}_{2}(1-\eta_2) \, z_{\rm D}\, \frac{ \partial F} { \partial z_ {\rm D}} + r_2\, u_2\, z_{\rm S}\, \frac{ \partial F} { \partial z_ {\rm S}}  }{\hat{N}_{\rm r, S, D}}     - \frac{  \tilde{r}_1\,(1-\eta_1)\, N_{\rm D}\, \frac{ \partial F} { \partial z_ {\rm D}} + r_1\, u_1\, N_{\rm S}\, \frac{ \partial F} { \partial z_ {\rm D}}  }  { \hat{N}_{\rm r, S, D}\, N_{\rm D} }   \right\}, \label{eq:generating}
\end{eqnarray}
where the operator
\begin{eqnarray}
\hat{N}_{\rm r,S,P}=r_{1}\,(N_{\rm S} -\hat{n}_{\rm S})  + r_{2}\, \hat{n}_{\rm S} + \tilde{r}_{1}(N_{\rm D}-\hat{n}_{\rm D}) +  \tilde{r}_{2}\, \hat{n}_{\rm D},
\end{eqnarray}
can be considered constant when $N_{\rm S}$ and $N_{\rm D}$ are set to be large, $ N_{\rm S}, N_{\rm D}\gg 1$.

Setting the coefficients of the derivatives $\frac{ \partial F} { \partial z_ {\rm S}} $ and $\frac{ \partial F} { \partial z_ {\rm D}} $ to zero, we obtain approximate quasi-stationary points of this constant population dynamics at the large--$t$ limit. This relates to the corresponding martingales for $N_{\rm S}, N_{\rm D} \to \infty$ branching process limit. Denoting the solutions with $z^{\star}_{\rm S}$ and $z^{\star}_{\rm P}$, one attains
\begin{eqnarray}
&& (z^{\star}_{\rm S}-1)\left[  r_2(1-u_2) \, z^{\star}_{\rm S}- r_1(1-u_1)-\tilde{r}_1\, \eta_1 \right] + (z^{\star}_{\rm D}-1)\, z_{\rm S}\, r_2\,u_2 =0 \nonumber\\[-3mm]
&& \label{eqbd111}  \\[-2mm]
&& (z^{\star}_{\rm D}-1)\left[ \tilde{r}_2\, (1-\eta_2 )\, z^{\star}_{\rm D} - \tilde{r}_1\,(1-\eta_1) - r_1\, u_1 \right] + (z^{\star}_{\rm S}-1)\, z^{\star}_{\rm D} \, \tilde{r}_2\, \eta_2=0. \nonumber
\end{eqnarray}
%
\subsection*{B. Fixation probability}

Taking advantage of the generating function, an exact analytic approach for the fixation probability can be derived even in the presence of plasticity (when mutation and mutation--back do not occur). The results are obtained for a BD Moran process; however, a similar calculation can be performed for a Voter (DB) Model with presumably different fixed points for the same initial conditions. The boundary and initial conditions for the corresponding generating function are resp. as follows
\begin{eqnarray}
&& F(z_{\rm S}=1,z_{\rm D}=1, t) = 1, \hspace{0.5cm} \mbox{for any $t > 0$},  \label{bound}\\
&& F(z_{\rm S}, z_{\rm D}, t = 0) = z_{\rm S}^{i}\, z_{\rm D}^{j}, \label{initial}
\end{eqnarray}
where $i$ and $j$ are the initial number of cancer SCs and DCs resp. For the following special cases, the system has two absorbing states at equilibrium which signify fixation and extinction for mutant cells in either compartments. We denote the probability of reaching extinction and fixation states by $B_0$ and $B_1$ resp. Thus we obtain the following result from the PGF at steady state
\begin{eqnarray}
F(z_{\rm S},z_{\rm D}, t \rightarrow \infty)  &=& p(n_{\rm S}=0, n_{\rm D}=0, t \rightarrow \infty)  \nonumber\\
&&  + p(n_{\rm S}= N_{\rm S}, n_{\rm d}=N_{\rm D}, t \rightarrow \infty) \,z_{\rm S}^{N_{\rm S}} \,z_{\rm D}^{N_{\rm D}} \nonumber\\
&= & B_0 + B_1\,z_{\rm S}^{N_{\rm S}} \,z_{\rm d}^{N_{\rm D}},
\end{eqnarray}
based on the boundary condition (\ref{bound}), $B_0+ B_1=1$. Biologically, since the mutant DCs are produced by cancer SCs, extinction of cancer SCs will result in replenishment of mutant DCs. Moreover, the co-operation between SCs and DCs suggests that it is impossible to have mutant SCs fixate while DCs become completely extinct. We conclude that
\begin{eqnarray}
F(z_{\rm S},z_{\rm D}, t \rightarrow \infty)  = 1- B_1 \left( 1- z_{\rm S}^{N_{\rm S}}\, z_{\rm D}^{N_{\rm D}}\right).
\end{eqnarray}

Finally, applying the initial condition, $F(z_{\rm S}, z_{\rm D}, t = 0) = z_{\rm S}^{i}\, z_{\rm D}^{j}$ and the boundary condition, $F(z_{\rm S}=1,z_{\rm D}=1, t) = 1$, the fixation probability for an initial population of $i$ malignant stem cells and $j$ progenitors is derived as
\begin{eqnarray}
\rho_{ij}= \frac{1- \left(z_{\rm S}^{\star}  \right)^i \,   \left(z_{\rm D}^{\star}  \right)^j }{  1- \left(z_{\rm S}^{\star}  \right)^{N_{\rm S}} \,\left(z_{\rm D}^{\star}  \right)^{N_{\rm D}} }, \label{FPP}
\end{eqnarray}
where $(z_{\rm S}^{\star},z_{\rm D}^{\star})$ is the nontrivial fixed point of the generating Eq. (\ref{eqbd111}).

\subsection*{C. Finding the quasi--fixed points and the associated survival probabilities}

An interesting scenario occurs when the normal component is not plastic, i.e. $\eta_{1}=0$, in which case the above equations can be simplified to the following expression of $z^{\star}_{\rm S},z^{\star}_{\rm D}\neq 1$
\begin{eqnarray}
z^{\star}_{\rm D} &=& \frac{  \Big( r_1\,(1-u_2)\,(\tilde{r}_1 + r_1\,u_1)  \Big) \,  z^{\star}_{\rm S} +  r_1\,(1-u_1)\,(\tilde{r}_1 + r_1\,u_1)    }{ \Big( r_1\,r_2\,(1-u_2)\,(1-\eta_1)-\tilde{r}_2\,r_2\, u_2\,\eta_2   \Big)\, z^{\star}_{\rm S}  - r_1\,r_2\,(1-u_1)\,(1-\eta_2) }.
\end{eqnarray}
Eqs. (\ref{eqbd}) can be reduced to the following closed form
\begin{eqnarray}
\left(  A - \frac{B}{z^{\star}_{\rm S}} \right) \left(  E- \frac{F}{z^{\star}_{\rm D}} \right)  = C\,.\,G,
\end{eqnarray}
where
\begin{eqnarray}
&& A= r_2\,(1-u_2),  \hspace{0.5cm}  B= r_1\,(1-u_1) + \tilde{r}_1\,\eta_1,  \hspace{0.5cm}  C= r_2\,u_2, \nonumber\\[-3mm]
&& \\[-2mm]
&& E= \tilde{r}_2\,(1-\eta_2),  \hspace{0.5cm}  F= \tilde{r}_1\,(1-\eta_1) + r_1\, u_1,  \hspace{0.5cm}  G= \tilde{r}_2\,\eta_2. \nonumber
\end{eqnarray}
The latter equations together with the original system suggest the following solution for $z_{\rm S}$ and thus represent another fixed point of the problem:
\begin{eqnarray}
(A^2 E -ACG)\, \left(z^{\star}_{\rm S} \right)^3 + ( ACF + C^2G + BCG + ACG -2 A B E - A^2 E -ACE )\, \left(z^{\star}_{\rm S} \right)^2 && \nonumber \\
 + ( B^2E + BCE + 2ABE - BCG -BCF  )\, z^{\star}_{\rm S}  = B^2 E.&& \label{eq999}
\end{eqnarray}

Now let us consider the following particular cases which give rise to some interesting consequences of the model.
\subsection*{C1. Standard Moran process}
Assume that there is no transition (migration) between SC and DC compartments (two islands), that is, $u_i, \eta_j \ll 1$ for $i,j=1,2$. Then each compartment will follow the mass--action BD Moran model with the fixed points for $z^{\star}_{\rm S}$ and $z^{\star}_{\rm D}$ where the overall behavior is akin to two disjoint Moran processes. The solutions to Eqs. (\ref{eqbd111}) are
\begin{itemize}
\item[(1)] $z^{\star}_{\rm S} = 1, \, z^{\star}_{\rm D} = 1,$
\item[(2)] $z^{\star}_{\rm S} = 1, \, z^{\star}_{\rm D} = \frac{\tilde{r}_1}{ \tilde{r}_2},$
\item[(3)] $z^{\star}_{\rm S} = \frac{r_1}{r_2}, \, z^{\star}_{\rm D} = 1,$
\item[(4)] $z^{\star}_{\rm S}=\frac{r_1}{r_2}, \, z^{\star}_{\rm D}=\frac{\tilde{r}_1}{ \tilde{r}_2}.$
\end{itemize}
where the solution (4) accounts for the fixed points of two separated Moran models in SC and Dc groups. Therefore, the fixation probabilities of starting from one imposed mutant in SC and DC groups resp. are
\begin{eqnarray}
\rho_{1}= \displaystyle{ \frac{1- \frac{r_1}{r_2} }{  1- \left(\frac{r_1}{r_2}  \right)^{N_{\rm S}}   }  },\hspace{1cm}
\rho_{2}=  \displaystyle{\frac{1- \frac{\tilde{r}_1}{ \tilde{r}_2} }{  1- \left(\frac{\tilde{r}_1}{ \tilde{r}_2}  \right)^{N_{\rm D}} }    }. \label{Noplast1}
\end{eqnarray}
%


Another interesting limit that results in a well-mixed Moran model in the SC class occurs when $\tilde{r}_{1,2} \simeq 0$ where SCs are also committed to reach a stationary state. Then the solutions to Eqs. (\ref{eqbd111}) are given by
\begin{itemize}
\item[(1)] $z^{\star}_{\rm S} = 1, \, z^{\star}_{\rm D} = 1,$
\item[(2)] $z^{\star}_{\rm S} = \frac{r_1}{r_2}, \, z^{\star}_{\rm D} = 1.$
\end{itemize}
Compared with the result from the previous case, the second solution relates to the solution of a well--mixed model occurring in the SC compartment where the only absorbing state for mutant DCs is to take over the whole population where the evolutionary scenario accounts for cancer DCs domination. In this scenario, the average fixation probability is
\begin{eqnarray}
\rho= {\displaystyle{ \frac{ N_{\rm S}\left(1-  \frac{r_1}{r_2}\right) }{ N_{\rm tot}\left(  1- \left(\frac{r_1}{r_2}  \right)^{N_{\rm S}}  \right) }  }},
\end{eqnarray}
%

\subsection*{C2. Invasion in hierarchical model (no plasticity)}

Now we consider the case where no plastic potential is taken into account during the whole  process, we assume that $r_1=\tilde{r}_1=1$, $r_2=\tilde{r}_2=r$, $u_1=u_2=u$, and $\eta_1=\eta_2=0$. in this situation, one obtains the following solutions
\begin{itemize}
\item[(1)] $z^{\star}_{\rm S} = 1, \, z^{\star}_{\rm D} = 1,$
\item[(2)] $z^{\star}_{\rm S} = \displaystyle{\frac{1}{u_{\mbox{eff}} }}, \, z^{\star}_{\rm D} = 1$
\item[(3)] $z^{\star}_{\rm S}= \frac{-r+u_1+u_2\,u_1-1+u_2 + \Gamma }{2r(-1+u_2)} , \, z^{\star}_{\rm D}=\frac{u_1+1}{ r}.$
\end{itemize}
where $u_{\mbox{eff}}=\frac{r(1-u_2)}{1-u_1}$ is the effective asymmetric division rate and
\begin{eqnarray}
\Gamma^2 &=& 1-6 r u_2 u_1+u_1^2+2 r u_2-2 r-2 u_2-2 u_1+2 r u_1+r^2+2 u_2 u_1^2+2 u_2^2 \nonumber\\
&& u_1+u_2^2+u_2^2 u_1^2.
\end{eqnarray}
The third solution cannot be maintained in reality since there is no cooperation between the compartments, which could result in support for the mutant SC by the mutant DC group. In such a case, the only possible outcome would be the fixation of the mutant DCs. Collectively, we find
\begin{eqnarray}
\rho_{\rm S}= \displaystyle{ \frac{1-  \frac{1-u_1}{r(1-u_2)} }{   1- \left(\frac{1-u_1}{r(1-u_2)}  \right)^{N_{\rm S}}}   }  ,
\hspace{0.5cm} \rho_{\rm D}= 0, \hspace{0.5cm} \rho=\displaystyle{ \frac{ N_{\rm S}}{N_{\rm tot}} }\,\rho_{\rm S}. \label{Heir1}
\end{eqnarray}

\subsection*{C3. Invasion with phenotypic plasticity (dedifferentiation)}
Suppose that dedifferentiation only occurs for the mutant DCs at a rate $\eta_2=\eta$ for  $\eta \gg \eta_1 \approx 0$. Also let assume that $r_1=\tilde{r}_1=1$, $r_2=\tilde{r}_2=r$, $u_1=u_2=u$ and $\eta_2=\eta$, then the solutions for $z^{\star}_{\rm S}$ and $z^{\star}_{\rm D}$ are
\begin{itemize}
\item[(1)] $z^{\star}_{\rm S} = 1, \, z^{\star}_{\rm D} = 1,$
\item[(2)] $z^{\star}_{\rm S}$ satisfies in the equation
\begin{eqnarray*}
&& \hspace{-1cm} (A^2 E -ACG)\, \left(z^{\star}_{\rm S} \right)^3 + ( ACF + C^2G + BCG + ACG -2 A B E - A^2 E -ACE )\, \left(z^{\star}_{\rm S} \right)^2 \nonumber \\
&&\hspace{-0.5cm} + ( B^2E + BCE + 2ABE - BCG -BCF  )\, z^{\star}_{\rm S}  = B^2 E,
\end{eqnarray*}
where $A=r(1-u), \, B=1-u, \, C=r u, \, E=r(1-\eta),\, F=u+1,\, G=r \eta$,
and $z^{\star}_{\rm D}=\frac{( z^{\star}_{\rm S} + 1)(1- u^2)}{ r[  (1-u-r u\eta) z^{\star}_{\rm S} - (1-u)(1-\eta) ]}.$
\end{itemize}
The values for $z^{\star}_{\rm S}$  and $z^{\star}_{\rm D}$, in this case, satisfy the following relations (for more details refer to Supplementary Information):
\begin{eqnarray}
&& \hspace{-1cm} \left(  A - \frac{B}{z^{\star}_{\rm S}} \right) \left(  E- \frac{F}{z^{\star}_{\rm D}} \right)  = C\,.\,G, \hspace{0.5cm} z^{\star}_{\rm D}=\frac{( z^{\star}_{\rm S} + 1)(1- u^2)}{ r[  (1-u-r u\eta) z^{\star}_{\rm S} - (1-u)(1-\eta) ]}, \label{eqgen1}
\end{eqnarray}
where $A=r(1-u), \, B=1-u, \, C=r u, \, E=r(1-\eta),\, F=u+1,\, G=r \eta$.

The derived solutions introduce the possible fixed points of the characteristic equations. We investigate this case in more detail later and through analyzing the phase diagram of the generalized model for non-zero plasticity in the replicator dynamics.

\subsection*{D. Phase diagram of the system}

Now, acquiring the replicator dynamics of the four compartment model which depict the alterations in the  the average frequencies of mutant SCs and DCs, resp. $x_{\rm S} = \langle n_{\rm S}(t)/N_{\rm S} \rangle$ and $x_{\rm D} = \langle n_{\rm D}/N_{\rm D} \rangle $, one obtains the following system of equations:
\begin{eqnarray}
\frac{d x_{\rm S}}{dt}   &\approx&    \left\langle  W^+_{{\rm S}} -W^-_{{\rm S}} \right\rangle    \nonumber\\
&=&  \frac{   [ r_2\,(1- u_2) - r_1\,(1-u_1)  ] \,x_{\rm S}\,(1- x_{\rm S})  + \tilde{r}_2\,\eta_2\,x_{\rm D}\,(1- x_{\rm S})  - \tilde{r}_1\eta_1\,x_{\rm S}\,(1- x_{\rm D})    } {  r_1\, (1-x_{\rm S}) + r_2\,  x_{\rm S} + \tilde{r}_1 \, (1-x_{\rm D}) +\tilde{r}_2 \, x_{\rm D} },  \nonumber \\[1mm]
&& \label{eq:RepM4}\\[-3mm]
\frac{d x_{\rm D}}{dt}   &\approx&      \left\langle  W^+_{{\rm D}} -W^-_{{\rm D}} \right\rangle  \nonumber \\
&=& \frac{   [ \tilde{r}_2\,(1- \eta_2) - \tilde{r}_1\,(1- \eta_1)  ] \,x_{\rm D}\,(1- x_{\rm D})  + r_2\,u_2\,x_{\rm S}\,(1- x_{\rm D})  - r_1\, u_1\,x_{\rm D}\,(1- x_{\rm S})    } {  r_1\, (1-x_{\rm S}) + r_2\,  x_{\rm S} + \tilde{r}_1 \, (1-x_{\rm D}) +\tilde{r}_2 \, x_{\rm D} }.  \nonumber
\end{eqnarray}

At equilibrium, the fraction of mutant stem and non-stem cell groups would lead to a specific states which are the pseudo--fixed points of the problem. These type of fixed points can be attractive or repulsive, depending on the initial conditions of the system. As we described in the paper, according to the cooperation between mutant SCs and DCs (and similarly between normal SCs and DCs), the malignant  individuals may become extinct or survive together. Now defining the fixed point as $(x^{\star}_{\rm S}, x^{\star}_{\rm D})$, it may tend either to $(0,0)$ or to $(1,1)$. Such a criteria suggests two distinct phases for the fate of the malignant cells which are separated by the phase boundary. At steady state, the following system can be derived for the fixed points $x^{\star}_{\rm S}$ and $x^{\star}_{\rm D}$:
\begin{eqnarray}
&&\hspace{-1.25cm} [ r_2\,(1- u_2) - r_1\,(1-u_1)  ] \,x^{\star}_{\rm S}\,(1- x^{\star}_{\rm S})  + \tilde{r}_2\,\eta_2\,x^{\star}_{\rm D}\,(1- x^{\star}_{\rm S})  - \tilde{r}_1\eta_1\,x^{\star}_{\rm S}\,(1- x^{\star}_{\rm D}) = 0,  \nonumber \\[-2mm]
&& \label{eqRD}\\[-2mm]
&&\hspace{-1.25cm} [ \tilde{r}_2\,(1- \eta_2) - \tilde{r}_1\,(1- \eta_1)  ] \,x^{\star}_{\rm D}\,(1- x^{\star}_{\rm D})  + r_2\,u_2\,x^{\star}_{\rm S}\,(1- x^{\star}_{\rm D})  - r_1\, u_1\,x^{\star}_{\rm D}\,(1- x^{\star}_{\rm S}) = 0.  \nonumber
\end{eqnarray}
To analyze the latter system, we may consider two important cases:
\paragraph{Case 1.} At first, we assume $u_1=u_2=\eta_1=\eta_2 =0$, then the solutions to the system (\ref{eqRD}) results in the extinction or fixation of mutant SCs and DCs. In this case, the phase diagram is simply divided to two advantageous ($r>1$) and disadvantageous ($r<1$) cases.

\paragraph{Case 2.}  Suppose that $r_1=\tilde{r}_1=1, r_2=\tilde{r}_2=r, u_1=u_2=u,  \eta_1= 0,$ and $\eta_2=:\eta$. This introduces an interesting scenario in which plasticity occurs only for the cancerous cells but not for the WT individuals. These restrictions simplify system (\ref{eqRD}) to the following possible cases for $x^{\star}_{\rm S}$ and $x^{\star}_{\rm D}$
\begin{itemize}
\item[(1)] $x^{\star}_{\rm S}=0, \hspace{0.3cm} x^{\star}_{\rm D} = 0$,
\item[(2)] $x^{\star}_{\rm S}=1, \hspace{0.3cm}  x^{\star}_{\rm D} = 1$,
\item[(3)] $x^{\star}_{\rm S}=1, \hspace{0.3cm} x^{\star}_{\rm D} = \frac{r\, u}{1-r+r\,\eta}$, and
\item[(4)] $x^{\star}_{\rm S}= r\,\eta\,{\cal A}\,{\cal M}^{-1}_1, \hspace{0.3cm} x^{\star}_{\rm D} = {\cal A}\,{\cal M}^{-1}_2,$
\end{itemize}
where
\begin{eqnarray}
&& {\cal M}_1 =3 r u^2+3 r+2 u-3 r^2 u^2+6 r^2 u-6 r u-r \eta-2 r^3 u-2 r^2 \eta u+r \eta u+2 r^2 \eta  \nonumber\\
&&\hspace{1cm}  +r^3 \eta u-3 r^2+r^3+r^3 u^2- r^3 \eta-1- u^2, \nonumber\\[-2mm]
&&  \label{MMA}\\[-2mm]
&& {\cal M}_2 = r^2 u+r^2 \eta- r^2-2 ru- r \eta+2 r- 1 +u, \nonumber\\
&&{\cal A}=2 r-1-r \eta+r^2 u+r^2 \eta+u^2-r u- r^2+r \eta u- r u^2.\nonumber
\end{eqnarray}

Among the given solutions (1)-(4) of the present case, the only acceptable non--trivial solution is (4). The solution (3) does not satisfy the condition $0\leq x_{\rm D}\leq 1$ for all possible values of $r,u,$ and $\eta$. Moreover, as we described above, the cooperation among mutant cells results in the fixation of both mutant SC and DC groups to the state $(1,1)$.

The last solution, then, implies that having $ (x_{\rm S},x_{\rm D})\rightarrow(0,0)$ leads to the following solution
\begin{eqnarray}
{\cal A}=(u+\eta-1)r^2+\left[-u^2+(\eta-1)u+2-\eta \right]r-1+u^2=0,
\end{eqnarray}
which characterizes the advantageous and disadvantageous regions for mutants (see Fig. (\ref{phase}) 
for more details). Another limit relates to the case where $ (x_{\rm S},x_{\rm D})$ approaches $(1,1)$. One obtains
\begin{eqnarray}
r\,\eta = (r-1)(u-1)
\end{eqnarray}
This condition, which is not defined for $r>1$, does not change the phase diagram and would not have any effect on the selection pressure of the system on mutants.

\bibliographystyle{spbasic}   
\bibliography{pheno_heterogeneity.bib}

\end{document}